\documentclass[fleqn,usenatbib]{mnras}

\usepackage{newtxtext,newtxmath}
\usepackage[T1]{fontenc}

\DeclareRobustCommand{\VAN}[3]{#2}
\let\VANthebibliography\thebibliography
\def\thebibliography{\DeclareRobustCommand{\VAN}[3]{##3}\VANthebibliography}

\usepackage{graphicx}	
\usepackage{amsmath,bm}	
\usepackage{caption}
\usepackage{subcaption}
\usepackage{hyperref}

\newcommand{\codename}{\textsc{jax-snana}}
\newcommand{\snana}{\textsc{snana}}
\newcommand{\jax}{\textsc{jax}}
\newcommand{\salt}{\textsc{salt}}
\newcommand{\bayesn}{\textsc{BayeSN}}

\newcommand{\saltReplayDdfCellY}{+0.41\,(0.74)}

\newcommand{\saltReplayDdfCellG}{+0.04\,(0.20)}

\newcommand{\saltReplayDdfCellI}{-0.01\,(0.25)}

\newcommand{\saltReplayDdfCellR}{-0.00\,(0.21)}

\newcommand{\saltReplayDdfCellU}{-0.03\,(0.32)}

\newcommand{\saltReplayDdfCellZ}{-0.12\,(0.28)}

\newcommand{\saltPopDdfNAccSnana}{33{,}132}
\newcommand{\saltPopDdfNAccJax}{33{,}486}
\newcommand{\saltPopDdfNGen}{100{,}000}

\newcommand{\saltPopDdfRejNepSnana}{29{,}671}
\newcommand{\saltPopDdfRejSearcheffSnana}{33{,}412}
\newcommand{\saltPopDdfRejCutwinSnana}{3{,}785}

\newcommand{\saltPopDdfRejNepJax}{29{,}469}
\newcommand{\saltPopDdfRejSearcheffJax}{33{,}254}
\newcommand{\saltPopDdfRejCutwinJax}{3{,}791}

\newcommand{\saltReplayWfdCellY}{+0.41\,(0.82)}

\newcommand{\saltReplayWfdCellG}{+0.03\,(0.19)}

\newcommand{\saltReplayWfdCellI}{-0.01\,(0.17)}

\newcommand{\saltReplayWfdCellR}{-0.00\,(0.16)}

\newcommand{\saltReplayWfdCellU}{+0.02\,(0.32)}

\newcommand{\saltReplayWfdCellZ}{-0.16\,(0.19)}

\newcommand{\saltPopWfdNAccSnana}{11{,}429}
\newcommand{\saltPopWfdNAccJax}{11{,}418}
\newcommand{\saltPopWfdNGen}{100{,}000}

\newcommand{\saltPopWfdRejNepSnana}{57{,}370}
\newcommand{\saltPopWfdRejSearcheffSnana}{27{,}854}
\newcommand{\saltPopWfdRejCutwinSnana}{3{,}347}

\newcommand{\saltPopWfdRejNepJax}{57{,}510}
\newcommand{\saltPopWfdRejSearcheffJax}{27{,}781}
\newcommand{\saltPopWfdRejCutwinJax}{3{,}291}

\newcommand{\bayesnReplayDdfCellY}{-0.05\,(1.57)}

\newcommand{\bayesnReplayDdfCellG}{-0.13\,(1.24)}

\newcommand{\bayesnReplayDdfCellI}{-0.40\,(1.22)}

\newcommand{\bayesnReplayDdfCellR}{-0.38\,(1.15)}

\newcommand{\bayesnReplayDdfCellU}{-0.38\,(1.97)}

\newcommand{\bayesnReplayDdfCellZ}{-0.52\,(1.09)}

\newcommand{\bayesnPopDdfNAccSnana}{34{,}440}
\newcommand{\bayesnPopDdfNAccJax}{34{,}506}
\newcommand{\bayesnPopDdfNGen}{100{,}000}

\newcommand{\bayesnPopDdfRejNepSnana}{30{,}132}
\newcommand{\bayesnPopDdfRejSearcheffSnana}{30{,}495}
\newcommand{\bayesnPopDdfRejCutwinSnana}{4{,}933}

\newcommand{\bayesnPopDdfRejNepJax}{30{,}135}
\newcommand{\bayesnPopDdfRejSearcheffJax}{30{,}579}
\newcommand{\bayesnPopDdfRejCutwinJax}{4{,}780}

\newcommand{\bayesnReplayWfdCellY}{-0.22\,(1.42)}

\newcommand{\bayesnReplayWfdCellG}{-0.15\,(1.22)}

\newcommand{\bayesnReplayWfdCellI}{-0.47\,(1.14)}

\newcommand{\bayesnReplayWfdCellR}{-0.46\,(1.09)}

\newcommand{\bayesnReplayWfdCellU}{-0.54\,(2.00)}

\newcommand{\bayesnReplayWfdCellZ}{-0.57\,(0.99)}

\newcommand{\bayesnPopWfdNAccSnana}{11{,}501}
\newcommand{\bayesnPopWfdNAccJax}{11{,}561}
\newcommand{\bayesnPopWfdNGen}{100{,}000}

\newcommand{\bayesnPopWfdRejNepSnana}{57{,}517}
\newcommand{\bayesnPopWfdRejSearcheffSnana}{26{,}502}
\newcommand{\bayesnPopWfdRejCutwinSnana}{4{,}480}

\newcommand{\bayesnPopWfdRejNepJax}{57{,}472}
\newcommand{\bayesnPopWfdRejSearcheffJax}{26{,}442}
\newcommand{\bayesnPopWfdRejCutwinJax}{4{,}525}

\newcommand{\contamReplaySniiCellY}{+0.26\,(1.12)}

\newcommand{\contamReplaySniiCellG}{-0.03\,(0.35)}

\newcommand{\contamReplaySniiCellI}{-0.06\,(0.97)}

\newcommand{\contamReplaySniiCellR}{-0.07\,(0.59)}

\newcommand{\contamReplaySniiCellU}{-0.10\,(0.59)}

\newcommand{\contamReplaySniiCellZ}{+0.00\,(1.45)}

\newcommand{\contamReplayNinetyOneBgCellY}{+0.50\,(2.29)}

\newcommand{\contamReplayNinetyOneBgCellG}{+0.10\,(1.12)}

\newcommand{\contamReplayNinetyOneBgCellI}{-0.28\,(2.51)}

\newcommand{\contamReplayNinetyOneBgCellR}{-0.08\,(1.81)}

\newcommand{\contamReplayNinetyOneBgCellU}{+0.80\,(1.78)}

\newcommand{\contamReplayNinetyOneBgCellZ}{-0.27\,(2.71)}

\newcommand{\contamReplayIaxCellY}{+0.47\,(3.51)}

\newcommand{\contamReplayIaxCellG}{-0.10\,(0.73)}

\newcommand{\contamReplayIaxCellI}{-0.64\,(2.80)}

\newcommand{\contamReplayIaxCellR}{-0.27\,(1.76)}

\newcommand{\contamReplayIaxCellU}{-0.07\,(0.61)}

\newcommand{\contamReplayIaxCellZ}{-0.13\,(4.05)}

\newcommand{\desReplayCellG}{+0.14\,(0.95)}

\newcommand{\desReplayCellI}{-0.16\,(0.37)}

\newcommand{\desReplayCellR}{-0.01\,(0.25)}

\newcommand{\desReplayCellZ}{-0.17\,(0.31)}

\newcommand{\desPopNAccSnana}{20{,}464}
\newcommand{\desPopNAccJax}{20{,}516}
\newcommand{\desPopNGen}{200{,}000}

\newcommand{\desPopRejNepSnana}{18{,}863}
\newcommand{\desPopRejSearcheffSnana}{160{,}148}
\newcommand{\desPopRejCutwinSnana}{525}

\newcommand{\desPopRejNepJax}{18{,}812}
\newcommand{\desPopRejSearcheffJax}{160{,}160}
\newcommand{\desPopRejCutwinJax}{512}

\newcommand{\fitSpeczNSalt}{1{,}680}

\newcommand{\fitSpeczSaltTzeroPullMedSci}{+5.0\times10^{-4}}
\newcommand{\fitSpeczSaltTzeroPullSigSci}{0.0097}

\newcommand{\fitSpeczSaltTzeroMadSnana}{0.57}
\newcommand{\fitSpeczSaltTzeroMadJax}{0.56}
\newcommand{\fitSpeczSaltTzeroSigRatio}{1.00}
\newcommand{\fitSpeczSaltTzeroTruthPullSnana}{0.77}
\newcommand{\fitSpeczSaltTzeroTruthPullJax}{0.77}

\newcommand{\fitSpeczSaltMBPullMedSci}{+0.0018}
\newcommand{\fitSpeczSaltMBPullSigSci}{0.0041}

\newcommand{\fitSpeczSaltMBMadSnana}{0.056}
\newcommand{\fitSpeczSaltMBMadJax}{0.056}
\newcommand{\fitSpeczSaltMBSigRatio}{1.00}
\newcommand{\fitSpeczSaltMBTruthPullSnana}{1.00}
\newcommand{\fitSpeczSaltMBTruthPullJax}{0.99}

\newcommand{\fitSpeczSaltXonePullMedSci}{+0.0015}
\newcommand{\fitSpeczSaltXonePullSigSci}{0.0033}

\newcommand{\fitSpeczSaltXoneMadSnana}{0.58}
\newcommand{\fitSpeczSaltXoneMadJax}{0.58}
\newcommand{\fitSpeczSaltXoneSigRatio}{1.00}
\newcommand{\fitSpeczSaltXoneTruthPullSnana}{1.00}
\newcommand{\fitSpeczSaltXoneTruthPullJax}{1.01}

\newcommand{\fitSpeczSaltCPullMedSci}{-0.0020}
\newcommand{\fitSpeczSaltCPullSigSci}{0.0046}

\newcommand{\fitSpeczSaltCMadSnana}{0.074}
\newcommand{\fitSpeczSaltCMadJax}{0.074}
\newcommand{\fitSpeczSaltCSigRatio}{1.00}
\newcommand{\fitSpeczSaltCTruthPullSnana}{1.25}
\newcommand{\fitSpeczSaltCTruthPullJax}{1.26}

\newcommand{\fitSpeczSaltNOutlier}{2}

\newcommand{\fitPhotozNopriorN}{762}

\newcommand{\fitPhotozNopriorZPullMedSci}{-0.0014}
\newcommand{\fitPhotozNopriorZPullSigSci}{0.039}

\newcommand{\fitPhotozNopriorZMadSnana}{0.15}
\newcommand{\fitPhotozNopriorZMadJax}{0.15}
\newcommand{\fitPhotozNopriorZSigRatio}{1.00}
\newcommand{\fitPhotozNopriorZTruthPullSnana}{3.01}
\newcommand{\fitPhotozNopriorZTruthPullJax}{3.25}

\newcommand{\fitPhotozNopriorTzeroPullMedSci}{-5.7\times10^{-4}}
\newcommand{\fitPhotozNopriorTzeroPullSigSci}{0.033}

\newcommand{\fitPhotozNopriorTzeroMadSnana}{2.3}
\newcommand{\fitPhotozNopriorTzeroMadJax}{2.3}
\newcommand{\fitPhotozNopriorTzeroSigRatio}{0.94}
\newcommand{\fitPhotozNopriorTzeroTruthPullSnana}{1.91}
\newcommand{\fitPhotozNopriorTzeroTruthPullJax}{1.86}

\newcommand{\fitPhotozNopriorMBPullMedSci}{-2.2\times10^{-4}}
\newcommand{\fitPhotozNopriorMBPullSigSci}{0.016}

\newcommand{\fitPhotozNopriorMBMadSnana}{0.13}
\newcommand{\fitPhotozNopriorMBMadJax}{0.13}
\newcommand{\fitPhotozNopriorMBSigRatio}{0.97}
\newcommand{\fitPhotozNopriorMBTruthPullSnana}{1.19}
\newcommand{\fitPhotozNopriorMBTruthPullJax}{1.32}

\newcommand{\fitPhotozNopriorXonePullMedSci}{-3.0\times10^{-7}}
\newcommand{\fitPhotozNopriorXonePullSigSci}{0.014}

\newcommand{\fitPhotozNopriorXoneMadSnana}{1.2}
\newcommand{\fitPhotozNopriorXoneMadJax}{1.1}
\newcommand{\fitPhotozNopriorXoneSigRatio}{0.98}
\newcommand{\fitPhotozNopriorXoneTruthPullSnana}{1.16}
\newcommand{\fitPhotozNopriorXoneTruthPullJax}{1.21}

\newcommand{\fitPhotozNopriorCPullMedSci}{+6.4\times10^{-4}}
\newcommand{\fitPhotozNopriorCPullSigSci}{0.024}

\newcommand{\fitPhotozNopriorCMadSnana}{0.21}
\newcommand{\fitPhotozNopriorCMadJax}{0.20}
\newcommand{\fitPhotozNopriorCSigRatio}{0.90}
\newcommand{\fitPhotozNopriorCTruthPullSnana}{1.87}
\newcommand{\fitPhotozNopriorCTruthPullJax}{2.33}

\newcommand{\fitPhotozGaussN}{878}

\newcommand{\fitPhotozGaussZPullMedSci}{+0.0012}
\newcommand{\fitPhotozGaussZPullSigSci}{0.031}

\newcommand{\fitPhotozGaussZMadSnana}{0.095}
\newcommand{\fitPhotozGaussZMadJax}{0.087}
\newcommand{\fitPhotozGaussZSigRatio}{1.00}
\newcommand{\fitPhotozGaussZTruthPullSnana}{2.47}
\newcommand{\fitPhotozGaussZTruthPullJax}{2.45}

\newcommand{\fitPhotozGaussTzeroPullMedSci}{+0.0016}
\newcommand{\fitPhotozGaussTzeroPullSigSci}{0.027}

\newcommand{\fitPhotozGaussTzeroMadSnana}{1.7}
\newcommand{\fitPhotozGaussTzeroMadJax}{1.7}
\newcommand{\fitPhotozGaussTzeroSigRatio}{0.95}
\newcommand{\fitPhotozGaussTzeroTruthPullSnana}{1.56}
\newcommand{\fitPhotozGaussTzeroTruthPullJax}{1.53}

\newcommand{\fitPhotozGaussMBPullMedSci}{-2.4\times10^{-4}}
\newcommand{\fitPhotozGaussMBPullSigSci}{0.012}

\newcommand{\fitPhotozGaussMBMadSnana}{0.11}
\newcommand{\fitPhotozGaussMBMadJax}{0.11}
\newcommand{\fitPhotozGaussMBSigRatio}{0.98}
\newcommand{\fitPhotozGaussMBTruthPullSnana}{1.17}
\newcommand{\fitPhotozGaussMBTruthPullJax}{1.17}

\newcommand{\fitPhotozGaussXonePullMedSci}{+2.9\times10^{-5}}
\newcommand{\fitPhotozGaussXonePullSigSci}{0.011}

\newcommand{\fitPhotozGaussXoneMadSnana}{1.1}
\newcommand{\fitPhotozGaussXoneMadJax}{1.1}
\newcommand{\fitPhotozGaussXoneSigRatio}{0.99}
\newcommand{\fitPhotozGaussXoneTruthPullSnana}{1.12}
\newcommand{\fitPhotozGaussXoneTruthPullJax}{1.10}

\newcommand{\fitPhotozGaussCPullMedSci}{-3.3\times10^{-4}}
\newcommand{\fitPhotozGaussCPullSigSci}{0.019}

\newcommand{\fitPhotozGaussCMadSnana}{0.15}
\newcommand{\fitPhotozGaussCMadJax}{0.15}
\newcommand{\fitPhotozGaussCSigRatio}{0.95}
\newcommand{\fitPhotozGaussCTruthPullSnana}{1.49}
\newcommand{\fitPhotozGaussCTruthPullJax}{1.70}

\newcommand{\fitPhotozQuantN}{836}

\newcommand{\fitPhotozQuantZPullMedSci}{+4.9\times10^{-4}}
\newcommand{\fitPhotozQuantZPullSigSci}{0.027}

\newcommand{\fitPhotozQuantZMadSnana}{0.086}
\newcommand{\fitPhotozQuantZMadJax}{0.087}
\newcommand{\fitPhotozQuantZSigRatio}{1.00}
\newcommand{\fitPhotozQuantZTruthPullSnana}{2.63}
\newcommand{\fitPhotozQuantZTruthPullJax}{2.77}

\newcommand{\fitPhotozQuantTzeroPullMedSci}{+0.0011}
\newcommand{\fitPhotozQuantTzeroPullSigSci}{0.027}

\newcommand{\fitPhotozQuantTzeroMadSnana}{1.7}
\newcommand{\fitPhotozQuantTzeroMadJax}{1.7}
\newcommand{\fitPhotozQuantTzeroSigRatio}{0.95}
\newcommand{\fitPhotozQuantTzeroTruthPullSnana}{1.53}
\newcommand{\fitPhotozQuantTzeroTruthPullJax}{1.56}

\newcommand{\fitPhotozQuantMBPullMedSci}{+3.1\times10^{-4}}
\newcommand{\fitPhotozQuantMBPullSigSci}{0.012}

\newcommand{\fitPhotozQuantMBMadSnana}{0.11}
\newcommand{\fitPhotozQuantMBMadJax}{0.11}
\newcommand{\fitPhotozQuantMBSigRatio}{0.99}
\newcommand{\fitPhotozQuantMBTruthPullSnana}{1.14}
\newcommand{\fitPhotozQuantMBTruthPullJax}{1.19}

\newcommand{\fitPhotozQuantXonePullMedSci}{+5.4\times10^{-4}}
\newcommand{\fitPhotozQuantXonePullSigSci}{0.011}

\newcommand{\fitPhotozQuantXoneMadSnana}{1.1}
\newcommand{\fitPhotozQuantXoneMadJax}{1.1}
\newcommand{\fitPhotozQuantXoneSigRatio}{0.99}
\newcommand{\fitPhotozQuantXoneTruthPullSnana}{1.09}
\newcommand{\fitPhotozQuantXoneTruthPullJax}{1.12}

\newcommand{\fitPhotozQuantCPullMedSci}{+3.2\times10^{-4}}
\newcommand{\fitPhotozQuantCPullSigSci}{0.016}

\newcommand{\fitPhotozQuantCMadSnana}{0.14}
\newcommand{\fitPhotozQuantCMadJax}{0.14}
\newcommand{\fitPhotozQuantCSigRatio}{0.96}
\newcommand{\fitPhotozQuantCTruthPullSnana}{1.50}
\newcommand{\fitPhotozQuantCTruthPullJax}{1.75}

\newcommand{\benchFitSaltN}{20{,}000}

\newcommand{\benchFitSaltSnana}{3.4}
\newcommand{\benchFitSaltSnanaCores}{110}
\newcommand{\benchFitSaltCpuSingle}{90}
\newcommand{\benchFitSaltCpuSingleSteady}{110}
\newcommand{\benchFitSaltGpuSingle}{1{,}500}
\newcommand{\benchFitSaltGpuSingleSteady}{5{,}000}
\newcommand{\benchFitSaltCpuDouble}{54}
\newcommand{\benchFitSaltCpuDoubleSteady}{66}
\newcommand{\benchFitSaltGpuDouble}{1{,}200}
\newcommand{\benchFitSaltGpuDoubleSteady}{2{,}800}
\newcommand{\benchFitSaltGpuN}{200{,}000}

\newcommand{\benchFitBayesnCpuSingle}{84}
\newcommand{\benchFitBayesnCpuSingleSteady}{90}
\newcommand{\benchFitBayesnGpuSingle}{2{,}900}
\newcommand{\benchFitBayesnGpuSingleSteady}{6{,}200}
\newcommand{\benchFitBayesnCpuDouble}{45}
\newcommand{\benchFitBayesnCpuDoubleSteady}{47}
\newcommand{\benchFitBayesnGpuDouble}{2{,}100}
\newcommand{\benchFitBayesnGpuDoubleSteady}{3{,}800}

\newcommand{\benchSimSaltDdfSnana}{100}
\newcommand{\benchSimSaltDdfSnanaCores}{3{,}300}
\newcommand{\benchSimSaltDdfCpuSingle}{3{,}900}
\newcommand{\benchSimSaltDdfCpuSingleSteady}{5{,}200}
\newcommand{\benchSimSaltDdfGpuSingle}{14{,}000}
\newcommand{\benchSimSaltDdfGpuSingleSteady}{56{,}000}
\newcommand{\benchSimSaltDdfCpuDouble}{3{,}400}
\newcommand{\benchSimSaltDdfCpuDoubleSteady}{4{,}300}
\newcommand{\benchSimSaltDdfGpuDouble}{13{,}000}
\newcommand{\benchSimSaltDdfGpuDoubleSteady}{41{,}000}

\newcommand{\benchSimBayesnDdfSnana}{190}
\newcommand{\benchSimBayesnDdfSnanaCores}{6{,}200}
\newcommand{\benchSimBayesnDdfCpuSingle}{2{,}100}
\newcommand{\benchSimBayesnDdfCpuSingleSteady}{2{,}400}
\newcommand{\benchSimBayesnDdfGpuSingle}{16{,}000}
\newcommand{\benchSimBayesnDdfGpuSingleSteady}{50{,}000}
\newcommand{\benchSimBayesnDdfCpuDouble}{1{,}500}
\newcommand{\benchSimBayesnDdfCpuDoubleSteady}{1{,}700}
\newcommand{\benchSimBayesnDdfGpuDouble}{13{,}000}
\newcommand{\benchSimBayesnDdfGpuDoubleSteady}{35{,}000}

\newcommand{\benchSimArnettDdfCpuSingle}{1{,}300}
\newcommand{\benchSimArnettDdfCpuSingleSteady}{3{,}400}
\newcommand{\benchSimArnettDdfGpuSingle}{10{,}000}
\newcommand{\benchSimArnettDdfGpuSingleSteady}{51{,}000}
\newcommand{\benchSimArnettDdfCpuDouble}{1{,}300}
\newcommand{\benchSimArnettDdfCpuDoubleSteady}{2{,}400}
\newcommand{\benchSimArnettDdfGpuDouble}{9{,}600}
\newcommand{\benchSimArnettDdfGpuDoubleSteady}{34{,}000}

\newcommand{\benchFitArnettDdfCpuSingle}{3.0}
\newcommand{\benchFitArnettDdfCpuSingleSteady}{3.1}
\newcommand{\benchFitArnettDdfGpuSingle}{260}
\newcommand{\benchFitArnettDdfGpuSingleSteady}{320}
\newcommand{\benchFitArnettDdfCpuDouble}{2.0}
\newcommand{\benchFitArnettDdfCpuDoubleSteady}{2.0}
\newcommand{\benchFitArnettDdfGpuDouble}{130}
\newcommand{\benchFitArnettDdfGpuDoubleSteady}{150}

\newcommand{\benchFitpzSaltDdfSnana}{0.67}
\newcommand{\benchFitpzSaltDdfSnanaCores}{21}
\newcommand{\benchFitpzSaltDdfCpuSingle}{16}
\newcommand{\benchFitpzSaltDdfCpuSingleSteady}{20}
\newcommand{\benchFitpzSaltDdfGpuSingle}{670}
\newcommand{\benchFitpzSaltDdfGpuSingleSteady}{1{,}100}
\newcommand{\benchFitpzSaltDdfCpuDouble}{11}
\newcommand{\benchFitpzSaltDdfCpuDoubleSteady}{12}
\newcommand{\benchFitpzSaltDdfGpuDouble}{340}
\newcommand{\benchFitpzSaltDdfGpuDoubleSteady}{480}

\newcommand{\benchFitpzBayesnDdfCpuSingle}{15}
\newcommand{\benchFitpzBayesnDdfCpuSingleSteady}{16}
\newcommand{\benchFitpzBayesnDdfGpuSingle}{960}
\newcommand{\benchFitpzBayesnDdfGpuSingleSteady}{1{,}300}
\newcommand{\benchFitpzBayesnDdfCpuDouble}{7.8}
\newcommand{\benchFitpzBayesnDdfCpuDoubleSteady}{8.1}
\newcommand{\benchFitpzBayesnDdfGpuDouble}{600}
\newcommand{\benchFitpzBayesnDdfGpuDoubleSteady}{720}

\newcommand{\benchSimSaltWfdSnana}{360}
\newcommand{\benchSimSaltWfdSnanaCores}{12{,}000}
\newcommand{\benchSimSaltWfdCpuSingle}{7{,}900}
\newcommand{\benchSimSaltWfdCpuSingleSteady}{12{,}000}
\newcommand{\benchSimSaltWfdGpuSingle}{25{,}000}
\newcommand{\benchSimSaltWfdGpuSingleSteady}{190{,}000}
\newcommand{\benchSimSaltWfdCpuDouble}{6{,}000}
\newcommand{\benchSimSaltWfdCpuDoubleSteady}{8{,}000}
\newcommand{\benchSimSaltWfdGpuDouble}{25{,}000}
\newcommand{\benchSimSaltWfdGpuDoubleSteady}{140{,}000}

\newcommand{\benchSimBayesnWfdSnana}{670}
\newcommand{\benchSimBayesnWfdSnanaCores}{21{,}000}
\newcommand{\benchSimBayesnWfdCpuSingle}{6{,}200}
\newcommand{\benchSimBayesnWfdCpuSingleSteady}{8{,}000}
\newcommand{\benchSimBayesnWfdGpuSingle}{31{,}000}
\newcommand{\benchSimBayesnWfdGpuSingleSteady}{170{,}000}
\newcommand{\benchSimBayesnWfdCpuDouble}{4{,}900}
\newcommand{\benchSimBayesnWfdCpuDoubleSteady}{5{,}900}
\newcommand{\benchSimBayesnWfdGpuDouble}{29{,}000}
\newcommand{\benchSimBayesnWfdGpuDoubleSteady}{130{,}000}

\newcommand{\benchSimArnettWfdCpuSingle}{840}
\newcommand{\benchSimArnettWfdCpuSingleSteady}{6{,}800}
\newcommand{\benchSimArnettWfdGpuSingle}{17{,}000}
\newcommand{\benchSimArnettWfdGpuSingleSteady}{250{,}000}
\newcommand{\benchSimArnettWfdCpuDouble}{810}
\newcommand{\benchSimArnettWfdCpuDoubleSteady}{5{,}800}
\newcommand{\benchSimArnettWfdGpuDouble}{18{,}000}
\newcommand{\benchSimArnettWfdGpuDoubleSteady}{160{,}000}

\newcommand{\benchFitArnettWfdCpuSingle}{4.1}
\newcommand{\benchFitArnettWfdCpuSingleSteady}{4.2}
\newcommand{\benchFitArnettWfdGpuSingle}{510}
\newcommand{\benchFitArnettWfdGpuSingleSteady}{730}
\newcommand{\benchFitArnettWfdCpuDouble}{3.1}
\newcommand{\benchFitArnettWfdCpuDoubleSteady}{3.1}
\newcommand{\benchFitArnettWfdGpuDouble}{330}
\newcommand{\benchFitArnettWfdGpuDoubleSteady}{410}

\newcommand{\benchFitpzSaltWfdSnana}{2.0}
\newcommand{\benchFitpzSaltWfdSnanaCores}{63}
\newcommand{\benchFitpzSaltWfdCpuSingle}{19}
\newcommand{\benchFitpzSaltWfdCpuSingleSteady}{27}
\newcommand{\benchFitpzSaltWfdGpuSingle}{900}
\newcommand{\benchFitpzSaltWfdGpuSingleSteady}{1{,}900}
\newcommand{\benchFitpzSaltWfdCpuDouble}{13}
\newcommand{\benchFitpzSaltWfdCpuDoubleSteady}{17}
\newcommand{\benchFitpzSaltWfdGpuDouble}{530}
\newcommand{\benchFitpzSaltWfdGpuDoubleSteady}{830}

\newcommand{\benchFitpzBayesnWfdCpuSingle}{50}
\newcommand{\benchFitpzBayesnWfdCpuSingleSteady}{64}
\newcommand{\benchFitpzBayesnWfdGpuSingle}{1{,}700}
\newcommand{\benchFitpzBayesnWfdGpuSingleSteady}{2{,}900}
\newcommand{\benchFitpzBayesnWfdCpuDouble}{27}
\newcommand{\benchFitpzBayesnWfdCpuDoubleSteady}{31}
\newcommand{\benchFitpzBayesnWfdGpuDouble}{1{,}200}
\newcommand{\benchFitpzBayesnWfdGpuDoubleSteady}{1{,}900}

\newcommand{\benchSimSaltN}{200{,}000}

\newcommand{\benchSimSaltGpuN}{1{,}000{,}000}

\newcommand{\benchSimArnettN}{20{,}000}

\newcommand{\benchSimArnettGpuN}{1{,}000{,}000}
\newcommand{\benchFitArnettN}{2{,}000}

\newcommand{\benchFitArnettGpuN}{100{,}000}
\newcommand{\benchFitpzSaltN}{5{,}000}

\newcommand{\benchFitpzSaltGpuN}{200{,}000}

\newcommand{\benchSnanaCoresPerGpu}{32}

\InputIfFileExists{generated/s08_bench_gpu.tex}{}{}
\newcommand{\bbcParitySpeczNReal}{50}

\newcommand{\bbcParitySpeczAlphaDMedSci}{+1.9\times10^{-6}}
\newcommand{\bbcParitySpeczAlphaDMadSci}{5.5\times10^{-6}}
\newcommand{\bbcParitySpeczAlphaSigRatioMed}{1.000}

\newcommand{\bbcParitySpeczBetaDMedSci}{+2.8\times10^{-5}}
\newcommand{\bbcParitySpeczBetaDMadSci}{3.8\times10^{-5}}
\newcommand{\bbcParitySpeczBetaSigRatioMed}{1.000}

\newcommand{\bbcParitySpeczGammaDMedSci}{+1.4\times10^{-6}}
\newcommand{\bbcParitySpeczGammaDMadSci}{4.2\times10^{-6}}
\newcommand{\bbcParitySpeczGammaSigRatioMed}{1.000}

\newcommand{\bbcParitySpeczOmDMedSci}{+5.7\times10^{-4}}
\newcommand{\bbcParitySpeczOmDMadSci}{1.6\times10^{-4}}
\newcommand{\bbcParitySpeczOmSigRatioMed}{0.998}

\newcommand{\bbcParitySpeczWDMedSci}{-1.3\times10^{-3}}
\newcommand{\bbcParitySpeczWDMadSci}{3.3\times10^{-4}}
\newcommand{\bbcParitySpeczWSigRatioMed}{1.001}

\newcommand{\bbcParityPhotNReal}{50}

\newcommand{\bbcParityPhotAlphaDMedSci}{+2.1\times10^{-6}}
\newcommand{\bbcParityPhotAlphaDMadSci}{6.7\times10^{-6}}
\newcommand{\bbcParityPhotAlphaSigRatioMed}{1.000}

\newcommand{\bbcParityPhotBetaDMedSci}{+5.9\times10^{-5}}
\newcommand{\bbcParityPhotBetaDMadSci}{7.7\times10^{-5}}
\newcommand{\bbcParityPhotBetaSigRatioMed}{1.000}

\newcommand{\bbcParityPhotGammaDMedSci}{+2.2\times10^{-5}}
\newcommand{\bbcParityPhotGammaDMadSci}{2.2\times10^{-5}}
\newcommand{\bbcParityPhotGammaSigRatioMed}{1.000}

\newcommand{\bbcParityPhotScalePCCDMedSci}{+4.0\times10^{-4}}
\newcommand{\bbcParityPhotScalePCCDMadSci}{7.1\times10^{-4}}
\newcommand{\bbcParityPhotScalePCCSigRatioMed}{0.984}

\newcommand{\bbcParityPhotOmDMedSci}{+4.5\times10^{-4}}
\newcommand{\bbcParityPhotOmDMadSci}{2.5\times10^{-4}}
\newcommand{\bbcParityPhotOmSigRatioMed}{0.999}

\newcommand{\bbcParityPhotWDMedSci}{-9.8\times10^{-4}}
\newcommand{\bbcParityPhotWDMadSci}{6.3\times10^{-4}}
\newcommand{\bbcParityPhotWSigRatioMed}{1.001}

\newcommand{\covParityNosysWJax}{-0.9966}
\newcommand{\covParityNosysWSnana}{-0.9966}
\newcommand{\covParityNosysSysJax}{0.0000}
\newcommand{\covParityNosysSysSnana}{0.0000}
\newcommand{\covParityBSTwentyOneWJax}{-0.9993}
\newcommand{\covParityBSTwentyOneWSnana}{-0.9993}
\newcommand{\covParityBSTwentyOneSysJax}{0.0059}
\newcommand{\covParityBSTwentyOneSysSnana}{0.0060}
\newcommand{\covParityDustOneWJax}{-0.9963}
\newcommand{\covParityDustOneWSnana}{-0.9963}
\newcommand{\covParityDustOneSysJax}{0.0018}
\newcommand{\covParityDustOneSysSnana}{0.0018}
\newcommand{\covParityDustTwoWJax}{-0.9954}
\newcommand{\covParityDustTwoWSnana}{-0.9954}
\newcommand{\covParityDustTwoSysJax}{0.0051}
\newcommand{\covParityDustTwoSysSnana}{0.0052}
\newcommand{\covParityDustThreeWJax}{-0.9958}
\newcommand{\covParityDustThreeWSnana}{-0.9958}
\newcommand{\covParityDustThreeSysJax}{0.0028}
\newcommand{\covParityDustThreeSysSnana}{0.0028}
\newcommand{\covParityWTwentyTwoWJax}{-0.9950}
\newcommand{\covParityWTwentyTwoWSnana}{-0.9950}
\newcommand{\covParityWTwentyTwoSysJax}{0.0035}
\newcommand{\covParityWTwentyTwoSysSnana}{0.0036}
\newcommand{\covParityMwebvWJax}{-0.9983}
\newcommand{\covParityMwebvWSnana}{-0.9983}
\newcommand{\covParityMwebvSysJax}{0.0025}
\newcommand{\covParityMwebvSysSnana}{0.0025}
\newcommand{\covParityCalSaltWJax}{-0.9896}
\newcommand{\covParityCalSaltWSnana}{-0.9896}
\newcommand{\covParityCalSaltSysJax}{0.0119}
\newcommand{\covParityCalSaltSysSnana}{0.0119}
\newcommand{\covParityCalspecWJax}{-0.9961}
\newcommand{\covParityCalspecWSnana}{-0.9961}
\newcommand{\covParityCalspecSysJax}{0.0019}
\newcommand{\covParityCalspecSysSnana}{0.0019}
\newcommand{\covParityZshiftWJax}{-0.9974}
\newcommand{\covParityZshiftWSnana}{-0.9974}
\newcommand{\covParityZshiftSysJax}{0.0032}
\newcommand{\covParityZshiftSysSnana}{0.0032}
\newcommand{\covParityMasslocWJax}{-0.9966}
\newcommand{\covParityMasslocWSnana}{-0.9966}
\newcommand{\covParityMasslocSysJax}{0.0006}
\newcommand{\covParityMasslocSysSnana}{0.0007}
\newcommand{\covParityAlphaevolWJax}{-0.9968}
\newcommand{\covParityAlphaevolWSnana}{-0.9968}
\newcommand{\covParityAlphaevolSysJax}{0.0000}
\newcommand{\covParityAlphaevolSysSnana}{0.0000}
\newcommand{\covParityBetaevolWJax}{-1.0002}
\newcommand{\covParityBetaevolWSnana}{-1.0002}
\newcommand{\covParityBetaevolSysJax}{0.0040}
\newcommand{\covParityBetaevolSysSnana}{0.0040}
\newcommand{\covParityGammaevolWJax}{-1.0040}
\newcommand{\covParityGammaevolWSnana}{-1.0040}
\newcommand{\covParityGammaevolSysJax}{0.0074}
\newcommand{\covParityGammaevolSysSnana}{0.0074}
\newcommand{\covParityCcpriorHElevenWJax}{-0.9966}
\newcommand{\covParityCcpriorHElevenWSnana}{-0.9966}
\newcommand{\covParityCcpriorHElevenSysJax}{0.0000}
\newcommand{\covParityCcpriorHElevenSysSnana}{0.0000}
\newcommand{\covParityColorlawWJax}{-0.9965}
\newcommand{\covParityColorlawWSnana}{-0.9965}
\newcommand{\covParityColorlawSysJax}{0.0027}
\newcommand{\covParityColorlawSysSnana}{0.0028}
\newcommand{\covParityFixabWJax}{-0.9966}
\newcommand{\covParityFixabWSnana}{-0.9966}
\newcommand{\covParityFixabSysJax}{0.0001}
\newcommand{\covParityFixabSysSnana}{0.0000}
\newcommand{\covParityIntrscColorWJax}{-0.9968}
\newcommand{\covParityIntrscColorWSnana}{-0.9968}
\newcommand{\covParityIntrscColorSysJax}{0.0000}
\newcommand{\covParityIntrscColorSysSnana}{0.0000}
\newcommand{\covParitySigintModelWJax}{-0.9962}
\newcommand{\covParitySigintModelWSnana}{-0.9962}
\newcommand{\covParitySigintModelSysJax}{0.0011}
\newcommand{\covParitySigintModelSysSnana}{0.0012}
\newcommand{\covParitySvahostWJax}{-0.9990}
\newcommand{\covParitySvahostWSnana}{-0.9990}
\newcommand{\covParitySvahostSysJax}{0.0020}
\newcommand{\covParitySvahostSysSnana}{0.0020}
\newcommand{\covParityEffZspecWJax}{-0.9961}
\newcommand{\covParityEffZspecWSnana}{-0.9961}
\newcommand{\covParityEffZspecSysJax}{0.0018}
\newcommand{\covParityEffZspecSysSnana}{0.0018}
\newcommand{\covParityAllWJax}{-1.0144}
\newcommand{\covParityAllWSnana}{-1.0144}
\newcommand{\covParityAllSysJax}{0.0213}
\newcommand{\covParityAllSysSnana}{0.0213}

\renewcommand{\benchSimArnettDdfCpuSingleSteady}{3{,}500}
\renewcommand{\benchSimArnettDdfCpuDoubleSteady}{2{,}400}
\renewcommand{\benchSimArnettDdfGpuSingleSteady}{410{,}000}
\renewcommand{\benchSimArnettDdfGpuDoubleSteady}{200{,}000}
\renewcommand{\benchSimArnettWfdCpuSingleSteady}{6{,}900}
\renewcommand{\benchSimArnettWfdCpuDoubleSteady}{5{,}900}
\renewcommand{\benchSimArnettWfdGpuSingleSteady}{1{,}200{,}000}
\renewcommand{\benchSimArnettWfdGpuDoubleSteady}{440{,}000}

\title[jax-SNANA: supernova simulation for the SBI era]{jax-SNANA: A GPU-accelerated Supernova Framework for the SBI Era}

\author[M. Grayling]{
M. Grayling$^1$\thanks{Email: mg2102@cam.ac.uk}
\\
$^1$ Institute of Astronomy and Kavli Institute for Cosmology, Madingley Road, Cambridge CB3 0HA, UK \\
}
\date{Accepted XXX. Received YYY; in original form ZZZ}
\pubyear{2026}

\begin{document}
\label{firstpage}
\pagerange{\pageref{firstpage}--\pageref{lastpage}}
\maketitle

\begin{abstract}
We present \codename{}, a \jax{} reimplementation of the supernova analysis framework \snana{}. \codename{} enables GPU-accelerated transient simulations and light curve fitting, supernova standardisation and cosmological inference. Running in single precision on an NVIDIA A100 GPU, \codename{} simulates more than 100,000 LSST-observed supernovae per second and fits 5,000 light curves per second after JIT compilation, a two orders-of-magnitude increase over current tools, essential for analysing samples at the scale expected from LSST. \codename{} currently supports the SALT and BayeSN spectral energy distribution (SED) models as well as physical models from \textsc{redback-jax} and a variety of SN SED templates; light curve fitting is supported with both photometric and spectroscopic redshifts. We fully validate the accuracy of \codename{} by comparing to \snana{}, finding excellent consistency between the two across a variety of tasks. The code provides a modular \jax{} end-to-end supernova cosmology pipeline, user-extensible to support new features including different SED templates and photometric classification algorithms. \codename{} has been designed with future applications of simulation-based inference in mind, providing fast simulations and enabling fully compiled on-the-fly simulations to be generated within the training loop of downstream SBI tasks. \codename{} extends beyond \snana{} by supporting simulations of generic SN-host galaxy relationships and multiple sub-populations of events, for example originating from young and old progenitors. \codename{} is publicly available at \url{https://github.com/mattgrayling/jax-snana} and installable through pip.
\end{abstract}

\begin{keywords}
methods: numerical -- software: simulations -- supernovae: general -- cosmology: observations
\end{keywords}


\section{Introduction}
\label{sec:intro}

Survey simulation is a vital tool across the field of transient astronomy. Simulations are used in many supernova cosmology analyses, characterising in fine detail the selection function of a survey and validating analysis pipelines \citep{Kessler17, Brout19}, and are often used to train photometric classification tools \citep[e.g.][]{Moller20}. Looking to the future, the field of simulation-based inference (SBI) is becoming increasingly prevalent across astronomy, a technique where simulations are used to learn the complex relationship between model parameters and observed data \citep{Cranmer20, Karchev23b}. In this work we present \codename{}, a full \jax{} reimplementation of the \snana{} framework supporting photometric and spectroscopic simulations, light curve fitting and full end-to-end cosmological inference \citep{Kessler09, jax}. Through GPU-acceleration enabled by \jax{}, \codename{} simulates tens of thousands of SNe/sec and fits thousands per second, transformational performance speed-ups which will prove vital for analyses at the scale of the Legacy Survey of Space and Time at Vera Rubin Observatory (LSST) \citep{Ivezic19} and pivotal for the development of SBI approaches.

The ability to create realistic survey simulations is essential for numerous aspects of time-domain astronomy, helping with both characterising the selection functions of existing surveys and forecasting the expected behaviour of future surveys. Unlike with static sources, the time-varying nature of these events adds extra complexity to realistic modelling of survey selection; for example, longer-lived transients with broader peaks and slower decline rates are more likely to be detected than those which exhibit fast declines, and the selected population is also sensitive to time-dependent effects such as observing conditions and weather. Accurate characterisation of survey selection functions is crucial for SN cosmology to obtain unbiased constraints on cosmological parameters and account for effects such as Malmquist bias \citep{Malmquist22}. The Beams with Bias Corrections (BBC) framework \citep{Kessler17} relies on accurate simulations to correct for selection bias, and has been used for a variety of recent SN cosmological analyses such as DES-SN5YR \citep{DES24, Vincenzi24}, Pantheon+ \citep{PantheonPlus, Brout22}, SN UNITE \citep{Camilleri26, Lee26} and Amalgame \citep{Popovic24Amalgame}. The UNITY supernova cosmology framework \citep{Rubin15, Rubin25, Rubin26} incorporates a treatment of selection effects directly within the inference, but still relies on simulations for validation. Accurate selection functions are also needed when calculating transient rates, and many rate analyses have used simulations to estimate detection efficiencies \citep[e.g.][]{Dilday10, Perrett12, Rodney14, Frohmaier19, Frohmaier21}. Disentangling intrinsic evolution with redshift of a population from changes in the observed population due to selection effects is an inherent challenge in the field. Simulations also play a key role in photometric classification; many classes of transient suffer from a lack of spectroscopically-confirmed examples which prevents photometric classification algorithms from training on real data. Instead, many are trained on large simulated datasets based on templates \citep[e.g.][]{Moller20, Muthukrishna19, Qu22}; a variety of photometric classification challenges, including PLAsTiCC and ELAsTiCC, have focused on training these algorithms on simulations \citep{Kessler19PLAsTiCC, ELAsTiCC23}. Many surveys will use realistic simulations to characterise the samples they expect to observe and calculate performance metrics, while simulations can also be used to explore and evaluate different observing strategies \citep[e.g.][]{Hounsell18, Rose21, Lochner22, Gris23}.

SBI is a technique gaining increased prevalence within transient astronomy, and astronomy in general \citep[e.g.][]{Cranmer20, Karchev23a}. The main strength of SBI is its ability to handle complex effects such as survey selection and photometric contamination \citep{Karchev23a, Boyd26}. As discussed, the selection functions of transient surveys are inherently complex and often there will be multiple selection functions at play; for example, a given object may require both good photometric coverage and a spectroscopic redshift (spec-$z$) of its host galaxy to be included in a cosmological sample. This means that the relationship between parameters of a given model and the observations corresponding to those parameters is often very complex itself; it is intractable to analytically express the likelihood for a model. SBI provides a natural solution to this problem, as neural networks can be used to learn how model parameters relate to observed data and in turn used for parameter inference \citep{Cranmer20}. For this reason, SBI is sometimes known as implicit-likelihood inference, as the likelihood is learned implicitly from simulations. There have been a variety of different SBI analyses of SNe~Ia \citep{Alsing18, Karchev23a, Boyd26, Karchev26, Popovic26} and other transients \citep{Villar22, Darc24, Desai25, Vidal25}, though in general the field is relatively new and is yet to be used for cosmological analysis on real data.

The \snana{} framework \citep{Kessler09} is a notable example of a simulation framework commonly used in time-domain astronomy. It has been used to forecast survey strategies, correct for selection bias in supernova cosmology, and generate transient-classification challenge datasets \citep{Hounsell18, Brout19, Kessler19PLAsTiCC}. \snana{} supports many features beyond simulations, however; as well as supporting simulations of a variety of different types of transients, it can be used for light curve fitting and provides a full end-to-end cosmological pipeline using the BEAMS with Bias Corrections method \citep{Kunz07, Hlozek12, Kessler17}. \snana{} is an incredibly detailed simulator, including many effects such as weak lensing, host-galaxy association and detection efficiency. However, in the era of LSST and SBI, \snana{} also faces a number of challenges. Many areas of astronomy are turning to GPU-accelerated simulators because of the huge performance increases they are able to achieve, which are essential for next-generation surveys \citep[e.g.][]{Li22, Piras23, Hahn24, Grayling24, Berry26}. SBI analyses are typically built using modern machine learning frameworks such as \jax{}, \textsc{pytorch} or \textsc{tensorflow} while \snana{} is largely written in C and Fortran, meaning it is not possible to directly interact with these frameworks or support GPU acceleration. The \textsc{python} package \textsc{sncosmo} supports simulations and light curve fitting, but not GPU acceleration and does not provide the same level of detail in simulations as \snana{} \citep{Barbary25}. Recent years have seen the development of newer simulators such as \textsc{skysurvey} \citep{Rigault26} and \textsc{LightCurveLynx} \citep{Dai26} with advantages over \snana{}, although these have focused purely on simulations rather than light curve fitting and cosmological inference.

For SBI approaches it is desirable to have fast simulators which can interact directly with the algorithms used for inference, which enables new training data to be simulated on the fly, though this is not essential. \snana{} has been used for some SBI analyses to date \citep{Popovic26, Boyd26}, though both relied on the creation of a large simulation prior to inference, either as a simulation bank to sample from or a large fixed sample to quantify the selection function. Other approaches have used bespoke simulations which do not incorporate the same level of realism as \snana{} simulations.

This work presents \codename{}, a GPU-accelerated \jax{} reimplementation of \snana{} which we make publicly available, providing support for simulations, light curve fitting and a full \jax{} end-to-end cosmological pipeline using BBC. This approach has three key advantages. First, \codename{} has been designed to optimise performance as much as possible for both CPU and GPU. On GPU, this code is capable of simulating tens of thousands of SNe/sec (post-JIT, it can simulate upwards of a hundred thousand SNe/sec) and also fitting thousands of light curves per second. Even running on a laptop, \codename{} provides a significant performance increase over \snana{}. The second advantage is the support for a fully compiled simulation pipeline which can interface directly with downstream SBI algorithms; \codename{} simulations can be included within the training loop and used to create on-the-fly training data with a fully compiled simulator held in GPU memory. The third advantage is that the code is fully implemented in \textsc{python} and designed for future extensibility by the community, supporting the easy addition of new models for simulating and fitting. 

\codename{} is designed to both match and extend beyond \snana{}, preserving compatibility with the same inputs while including additional functionality. At present \codename{} supports the SALT \citep{Guy07, Kenworthy21} and BayeSN \citep{Mandel22, Grayling24} SN~Ia SED models, as well as the spectral templates for core-collapse SNe and peculiar SNe~Ia used within \snana{} \citep{Jones17, Vincenzi19, Kessler19PLAsTiCC}. Through \textsc{redback-jax} (\citealt{Sarin24}; \textcolor{blue}{Sarin et al., in prep.}), it also supports more physically-motivated models such as the Arnett model \citep{Arnett82, Pinto00, Sarin26b}, magnetar-powered supernovae \citep{Nicholl17, Guillochon18, Omand24, Sarin22}, tidal disruption events \citep{Guillochon13, Mockler19}, shock cooling \citep{Piro21}, shocked cocoon emission \citep{Piro18}, interaction with circumstellar material \citep{Sarin26c} and kilonovae \citep{Metzger17}.

The paper is structured as follows. Section~\ref{sec:framework} details the framework itself, including how it was developed and the structure of different elements such as simulation, light curve fitting and cosmological inference. Section~\ref{sec:validation} presents a comparison with outputs from \snana{} using the same inputs to validate \codename{} and demonstrate its accuracy. Section~\ref{sec:performance} provides performance benchmarks on both CPU and GPU to demonstrate the power of this framework to scale to next-generation data samples. Section~\ref{sec:discussion} discusses the benefits \codename{} can have for future analyses with LSST and Roman, and Section~\ref{sec:conclusions} presents our conclusions.

\section{The \codename{} framework}
\label{sec:framework}

This section outlines the \codename{} framework itself, including some of the design decisions that were made. The development of \codename{} was enabled by the use of agentic coding tools, namely Claude Code by Anthropic and Codex by OpenAI, working under close supervision with human-led architecture and design decisions. Overall, \codename{} has been designed to be fully modular, allowing users to easily add new functionality such as additional SED models and templates or new photometric classification algorithms.

\subsection{Supported models}

At release, \codename{} supports SALT2 and SALT3 SN~Ia models \citep{Guy07, Guy10, Kenworthy21, Taylor23, PopovicDovekie}, the BayeSN model \citep{Mandel22, T21, Grayling24, Grayling26} and templates of core-collapse supernovae and peculiar SNe~Ia from \citet{Jones17}, \citet{Vincenzi19} and the PLAsTiCC model library \citep{Kessler19PLAsTiCC}. Support is also included for \textsc{redback} models \citep{Sarin24} through \textsc{redback-jax} (\textcolor{blue}{Sarin et al., in prep.}), a \jax{} port of \textsc{redback}, allowing for realistic survey simulations with a variety of physically-motivated models including the Arnett radioactive decay model \citep{Arnett82, Pinto00, Sarin26b}, magnetar models \citep{Nicholl17, Guillochon18, Omand24, Sarin22}, shock cooling models \citep{Piro21}, shocked cocoon emission \citep{Piro18}, interaction with circumstellar material \citep{Sarin26c}, tidal disruption events \citep{Guillochon13, Mockler19} and kilonovae \citep{Metzger17}. All of these models are available for simulation. Light curve fitting currently supports SALT, BayeSN, and \textsc{redback-jax} models.

\subsection{Designing \codename{}}
\label{sec:design}

Much of the computational structure of \codename{} was inspired by the GPU-accelerated \jax{} BayeSN framework presented in \citet{Grayling24}; this was structured carefully to ensure that all model SED evaluation and photometry was structured as a series of fixed-shape tensor operations, allowing it to run very efficiently on GPU and make maximum use of fast compiled code. However, in \codename{} great care had to be taken to maximise performance. \jax{} code can provide huge performance speed-ups over other \textsc{python} packages because of its use of Just-in-time (JIT) compilation; the first time a given operation of a given size is run, the code is compiled to run very efficiently on the device being used (where available, a GPU). As a result, the first time the code is called there is a large compilation overhead, whereas the second time the code will run near-instantaneously. For light curve fitting, where the same function is called repeatedly as part of inference or optimisation, the overall runtime will be considerably reduced despite the compilation overhead the first time the code is called. However, for simulation each SN only requires one pass of the code, meaning that with a naive implementation JIT-compilation would add overhead without any benefit. The solution to this was to simulate SNe in batches of the same size; the first batch pays compilation overhead, but each subsequent batch is simulated incredibly quickly. This approach means that larger simulations become increasingly efficient, as the compilation overhead remains the same and is effectively spread out over more total simulated SNe. This requires each batch to contain the same number of SNe and the same total number of data points, not trivial considering that different SNe have different numbers of observations. This problem was circumvented by padding each SN to the length of the maximum possible number of data points covered by the model being simulated over the observing log being used to simulate realistic observations. While this does slightly increase memory footprint, this is more than counteracted by the huge performance increase resulting from preserving a constant operation size over all batches and avoiding recompilation.

In a number of cases methods used by \snana{} had to be adapted to be suitable for vectorisation, as \snana{} was written with serial code in mind. For example, \snana{} used rejection sampling in many cases to apply truncation to distributions. \codename{} deliberately avoids rejection sampling as it is inefficient for vectorised compute which requires fixed numbers of samples. Instead, the inverse-CDF method is used routinely to enable sampling of parameters from arbitrary distributions.

\subsection{Simulation}
\label{sec:sim}

In this section we present details of the simulation pipeline in \codename{} and give examples of simulated light curves and spectra for a variety of models.

\subsubsection{Simulation pipeline}

\begin{figure}
	\includegraphics[width=\columnwidth]{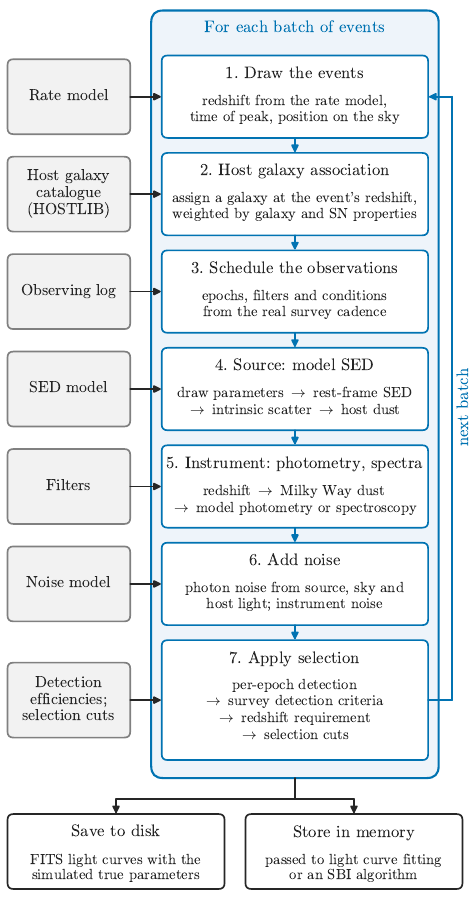}
    \caption{This flowchart illustrates the full \codename{} simulation pipeline.}
    \label{fig:sim_flowchart}
\end{figure}

Fig. \ref{fig:sim_flowchart} shows a flowchart illustrating the simulation pipeline used by \snana{} and \codename{}. This involves several stages:

\begin{enumerate}
    \item Properties of each transient are drawn. This includes its cosmological redshift, peculiar velocity and time-of-maximum.
    \item The transient is optionally associated with a host galaxy, as discussed in Section~\ref{sec:host_assocation}.
    \item Observations are scheduled for a given SN, based upon observations of the field it is located in from the observing log, its time-of-maximum and the phase range covered by the transient SED model being simulated.
    \item The model SED is evaluated for the transient based on the SED in question and the drawn parameters.
    \item The model SED is used to evaluate noiseless photometry and/or spectra, depending on what is requested.
    \item Noise is applied to the observations, based on the observing conditions in the observing log, photon noise from the source and light from the host galaxy. 
    \item Selection effects are then applied to the simulated observations, to recreate the selection cuts applied in real data analyses.
\end{enumerate}

As discussed in Section \ref{sec:design} this full pipeline is JIT-compiled meaning that batches of SNe can be called repeatedly very efficiently.

\subsubsection{Treatment of peculiar velocities}

Peculiar velocities change both the observed redshift and the observed brightness of a supernova. Following \citet{Davis11} and \citet{Carr22}, the observed heliocentric redshift, $z_{\rm hel}$, is related to the cosmological redshift, $z_{\rm HD}$, by
\begin{equation}
    1 + z_{\rm hel} = (1 + z_{\rm HD})(1 + z_{\rm Sun})(1 + z_{\rm p}),
\end{equation}
where $z_{\rm Sun}$ is the redshift due to our own motion relative to the Cosmic Microwave Background (CMB) and $z_{\rm p} \simeq v_{\rm p}/c$ that due to the peculiar velocity of the supernova. In the absence of peculiar velocities, the luminosity distance is $d_L = (1 + z_{\rm HD})\,D_M(z_{\rm HD})$, where $D_M$ is the transverse comoving distance. Peculiar motions also change the observed flux: our own motion contributes one factor of $(1 + z_{\rm Sun})$, and the motion of the supernova contributes two factors of $(1 + z_{\rm p})$, one from the Doppler shift of the photons and one from relativistic beaming \citep{Davis11, Carr22}. The luminosity distance is therefore
\begin{align}
    d_L &= (1 + z_{\rm HD})\,D_M(z_{\rm HD})\,(1 + z_{\rm Sun})(1 + z_{\rm p})^2 \nonumber \\
        &= (1 + z_{\rm hel})\,D_M(z_{\rm HD})\,(1 + z_{\rm p}).
\end{align}
Both \snana{} and \codename{} compute the distance modulus, $\mu = 5\log_{10}(d_L/10\,{\rm pc})$, in this way, integrating over the cosmological redshift and using the heliocentric redshift in the prefactor. Peculiar velocities can be drawn from a Gaussian distribution, fixed to a single value, or taken for each host galaxy from a column of the host-galaxy library. This approach is notably different to \textsc{python} cosmology libraries such as \textsc{jax-cosmo} \citep{Campagne23} and \textsc{astropy} \citep{Astropy13, Astropy18, Astropy22}, which compute distances from a single redshift (or scale factor) and do not distinguish between the cosmological and heliocentric redshifts, or include the peculiar-velocity terms.

\subsubsection{Host galaxy association}
\label{sec:host_assocation}

\codename{} follows the same approach for associating host galaxies with SNe as \snana{}. This involves a large catalogue of galaxies (known as a HOSTLIB in \snana{}; \citealt{Gupta16}) with known properties such as photometry, host galaxy stellar mass, star formation rate and light-profile (S\'ersic) parameters used to place the SN within its host; these can also have corresponding photometric redshift (photo-$z$) distributions based on their photometry. After properties are drawn for a given transient, it is assigned to a host galaxy close to the simulated redshift of the transient. Optionally, a separate file containing weights (known in \snana{} as a WGTMAP) can be provided to modify the probability of SNe being assigned to different types of galaxies depending on their properties. For example, this approach can be used to simulate the well-established link between SN~Ia light curve shape and the properties of its environment \citep{Sullivan10, Rigault20, Ginolin25}.

This approach relies on the assumption of a complete unbiased galaxy catalogue for host association. Other approaches have been followed by other simulators; for example, \textsc{skysurvey} \citep{Rigault26} first draws host galaxy properties and then draws the properties of the supernovae associated with that host, which allows for exploration of a variety of links between transient properties and host galaxy properties. This approach of forward modelling the full galaxy population \citep[see also][]{Thorp25} avoids the need for a complete catalogue, and this alternate path could be explored for \codename{} in future.

\subsubsection{Simulating sub-populations and SN-host connections}
\label{sec:sn_host}

While the primary goal of \codename{} has been to reimplement \snana{} in \jax{}, there are a number of cases where extra features have been added. An example of this is in the treatment of the link between SN and host properties. There are many established links between the properties of SNe and the properties of their host environments, for example core-collapse supernovae are associated with star forming regions due to their massive star progenitors \citep[e.g.][]{Anderson12, Grayling23}, and superluminous supernovae are associated with low-mass, low-metallicity hosts \citep[e.g.][]{Lunnan14, Perley16}. It has long been established that there is a link between the properties of SNe~Ia and their host galaxies \citep{Kelly10, Sullivan10}, and there is extensive research into the exact cause of this \citep[e.g.][]{Rigault13, Rigault20, Childress13, Childress14, Roman18, Jones18, Rose19, Kelsey21, Kelsey23, Wiseman22, Wiseman23, Briday22, Meldorf23, BS21, Popovic21, Popovic23, Nicolas21, Wojtak23, Thorp24, TM22, Grayling24, Popovic24Dust, GraylingPopovic25, Ginolin25, Ginolin25b, Popovic25Dust, Ginolin26, Kelsey26}. 

\snana{} allows for simulated SNe to be associated preferentially with specific types of hosts using a WGTMAP file, as discussed in Section~\ref{sec:host_assocation}. While we support this functionality in \codename{}, we also allow for the simulation of generic relations between SN properties and host properties directly as specified by the user without requiring a separately-built WGTMAP file, allowing maximum customisability to explore the impact of different SN-host correlations on observations. 

Previous work looking at the environmental-dependence of SNe~Ia has also suggested that there may be two separate populations of events; a fast-declining mode associated with old progenitors and a slower-declining mode associated with younger progenitors \citep[e.g.][]{Rigault20, Nicolas21, Wiseman22, Larison24, Ginolin25}. This would explain established links between host-mass, redshift and light curve stretch, and could also help to explain the `mass-step' (the offset in standardised SN~Ia brightness between SNe in high- and low-mass host galaxies; \citealt{Kelly10, Sullivan10, Lampeitl10}). Motivated by this, \codename{} enables the simulation of multiple sub-populations with different parameter distributions and an optional luminosity offset between them. Membership of these sub-populations can be linked to host properties such as specific star formation rate, as a proxy for age.

\begin{figure}
	\includegraphics[width=\columnwidth]{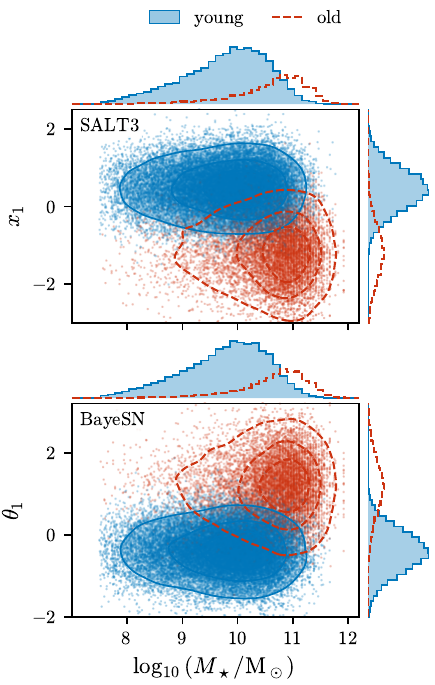}
	\caption{\textbf{Top:} Simulated SALT $x_1$ distributions for an example \codename{} simulation of two SN~Ia populations with different stretch distributions associated with old and young environments. This simulation uses the COSMOS2020 \citep{Weaver22} sample as a host galaxy library, with supernovae assigned to the old and young classes based on the star formation rate and mass of the host. This illustrative, simplified simulation demonstrates \codename{} can be used to simulate various relations between SN properties and galaxy properties. \textbf{Bottom:} As above but with BayeSN simulations, using $\theta_1$ as a stretch parameter.}
	\label{fig:two_population}
\end{figure}

\begin{figure}
	\includegraphics[width=\columnwidth]{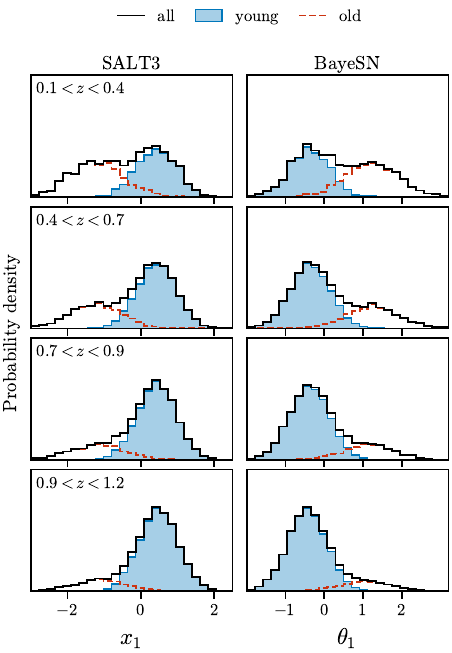}
	\caption{Stretch parameter distributions for SALT and BayeSN across different redshift bins for the simulation presented in Fig.~\ref{fig:two_population}. Note that no redshift evolution was specifically encoded in the simulation; instead the relative balance of the two populations naturally evolves as the galaxy population evolves.}
	\label{fig:two_population_z}
\end{figure}

Figs.~\ref{fig:two_population} and~\ref{fig:two_population_z} present an illustrative example of how this works in practice. This simple simulated model aims to reproduce the observed bimodality in SN~Ia stretch distributions at low redshift, but not at high redshift \citep{Nicolas21, Ginolin25}. We consider both a SALT simulation, with $x_1$ as the stretch parameter, and a BayeSN simulation with $\theta_1$ as the stretch parameter. These are LSST simulations of the Deep Drilling Fields (DDFs), using simulated observing strategies \citep{Naghib19, Bianco22, Biswas20}. Here, we simulate a young and an old population of SNe whose relative rates follow the star formation rate and stellar mass of the host galaxy \citep{Sullivan06}, respectively drawn from the high- and low-stretch modes \citep{Ginolin25}. We simulate using the COSMOS2020 \citep{Weaver22} galaxy population as a host library; although we do not explicitly include any redshift evolution in the model, the evolution in host properties across the sample with redshift leads to the expected result where the stretch distribution is near-unimodal at higher redshifts but becomes increasingly bimodal at lower redshifts; we also observe a clear link between SN stretch and host galaxy stellar mass, as expected \citep[e.g.][]{Sullivan10, Childress13, Ginolin25}.

\subsubsection{Detection efficiency pipeline}
\label{sec:searcheff}

\snana{}, and hence \codename{}, aim to recreate the selection function of a real survey. There are a number of stages to recreate the inherently random process of source detection, outlined as follows:

\begin{enumerate}
    \item There is initially a detection efficiency pipeline, which gives the probability that the difference-imaging pipeline would have detected a source of that signal-to-noise, with each individual observation randomly flagged as detected or not accordingly.
    \item Following the per-observation detection flag, there is an object-level pipeline which considers all individual detections of an event to determine if the transient itself would have been discovered. This can depend on the exact detection criteria used; for example, a transient source may be considered as detected based on the presence of two detections in different filters some minimum time period apart.
    \item Optionally, there is a follow-up stage which models whether each individual discovered event would receive spectroscopic follow-up; this can model the probability of each transient receiving a classification spectrum in the case of a spectroscopically-confirmed sample, and a host galaxy spectrum in the case of a sample with spec-$z$s. In the former case this is parametrised as a function of SN properties such as peak magnitude, and potentially light curve shape and colour. For the latter this instead depends on the magnitude of the host galaxy \citep[e.g.][]{Vincenzi21, Chen26}.
\end{enumerate}

Through this process \codename{} simulates the process of detecting new transients and obtaining follow-up to model realistic survey selection. The follow-up stage is optional and can be omitted to model a purely photometric survey. The pipeline efficiencies used as input for these stages are typically measured by injecting fake sources into real survey images to ensure this captures the real selection function \citep[e.g.][]{Kessler15, Vincenzi21}.

In \codename{} each of these stages is evaluated for a whole batch of simulated events at once, with the outcomes recorded to allow the results to be directly compared to \snana{} (see Section \ref{sec:validation}). Following this process, data-specific quality cuts are also applied based on e.g. the signal-to-noise of the observations, the number of epochs and phase coverage, recreating the full selection process of a real sample.

\subsubsection{Example simulations}

\begin{figure}
	\includegraphics[width=\columnwidth]{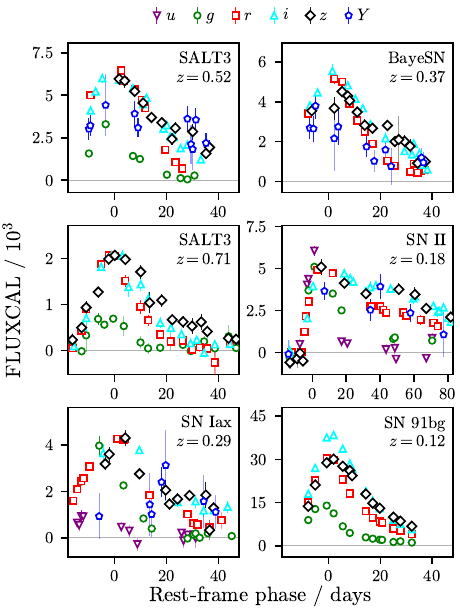}
    \caption{Example \codename{}-simulated LSST and DES-SN5YR light curves for a variety of different SN SED models and templates.}
    \label{fig:sim_lightcurves}
\end{figure}

\begin{figure}
	\includegraphics[width=\columnwidth]{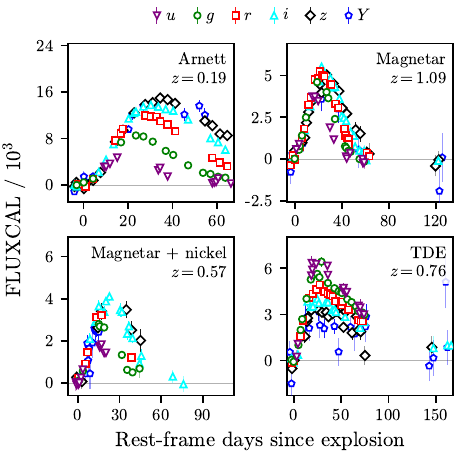}
    \caption{Example simulated LSST DDF light curves using various \textsc{redback-jax} models (\citealt{Sarin24}; \textcolor{blue}{Sarin et al., in prep.}): the Arnett nickel-decay model (\texttt{arnett}; \citealt{Arnett82, Pinto00, Sarin26b}), the magnetar spin-down model (\texttt{magnetar\_powered}; \citealt{Nicholl17, Sarin22}), nickel decay plus a magnetar (\texttt{magnetar\_nickel}; \citealt{Guillochon18}), and the tidal-disruption fallback model (\texttt{tde\_fallback}; \citealt{Guillochon13, Mockler19, Guillochon18}).}
    \label{fig:redback_lightcurves}
\end{figure}

\begin{figure}
	\includegraphics[width=\columnwidth]{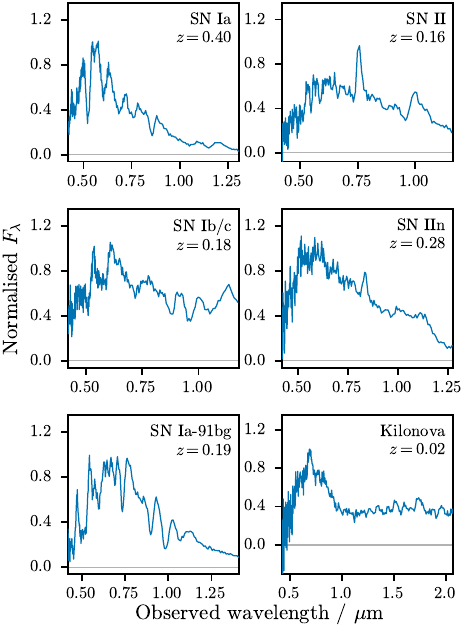}
    \caption{Example simulated spectra, using the Roman IFU spectrograph implementation from \citet{Hounsell18}, for SALT3 \citep{Kenworthy21}, SN II, SN IIn and SN Ib/c templates from \citet{Vincenzi19}, the SN~Ia-91bg templates from the PLAsTiCC model library \citep{Kessler19PLAsTiCC} and a binary neutron star kilonova model computed with POSSIS \citep{Bulla19}.}
    \label{fig:spectra}
\end{figure}

We now illustrate some examples of data simulated using \codename{}. First, Fig. \ref{fig:sim_lightcurves} showcases some examples of different simulated light curves using a variety of SED models and templates including SALT3, BayeSN and a variety of templates from \citet{Vincenzi19} and the PLAsTiCC model library \citep{Kessler19PLAsTiCC}. Fig. \ref{fig:redback_lightcurves}, meanwhile, shows simulated light curves from a variety of models implemented in \textsc{redback-jax} (\textcolor{blue}{Sarin et al., in prep.}). Finally, Fig. \ref{fig:spectra} shows examples of simulated spectra, using a model of the Roman IFU spectrograph from \citet{Hounsell18}.

\subsection{Light curve fitting}
\label{sec:fit}

\begin{figure}
	\includegraphics[width=\columnwidth]{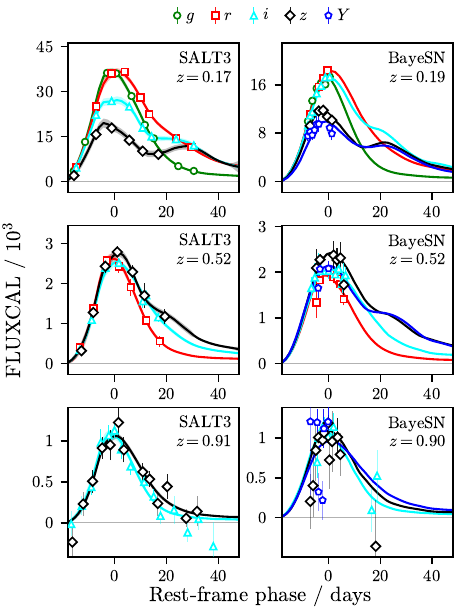}
    \caption{Example \codename{} light curve fits to simulated DES light curves with SALT (left panels) and simulated LSST light curves with BayeSN (right panels).}
    \label{fig:fit_lightcurves}
\end{figure}

\codename{} provides support for light curve fitting using the SALT and BayeSN SED models, and Fig.~\ref{fig:fit_lightcurves} presents some example fits to simulated DES and LSST light curves. Fitting is particularly well suited for \jax{} and JIT-compilation because of the repeated calls of the same function. In addition, \jax{} provides gradients through autodiff \citep{jax} making it ideal for gradient-based exploration of parameter spaces. However, there were also complications associated with light curve fitting. In general GPUs are able to provide a significant performance increase through efficient tensor operations for large vectorised computations. This works straightforwardly for supernova simulations as the SED model construction and photometry can be vectorised, enabling very efficient simulations. For light curve fitting, each SN must have its own independent optimiser while preserving the vectorised computation of model photometry across all SNe at once. There are a number of existing packages for optimisation implemented in \jax{}, including \textsc{jaxopt} \citep{Blondel21}; however, for \codename{} the best results and performance were found using a custom batched implementation of the Levenberg-Marquardt optimiser \citep{Levenberg44, Marquardt63} which uses gradients and Hessians calculated through autodiff. This allows independent events to be fit together on GPU.

\codename{} implements the multi-stage approach to light curve fitting employed by \snana{}, described as follows:

\begin{enumerate}
    \item There is an initial coarse grid search to identify parameter values to start the optimiser at, mostly focused on peak MJD.
    \item Following the grid search, there is a first round of optimisation. The SALT SED model includes an error term, which is a function of the SN parameters being fit for, that is included in the denominator of the loss function; in the first fitting round this error is recomputed for every proposed set of parameter values. BayeSN does not include such an error model, but the intrinsic scatter term $\epsilon$ (see \citealp{Mandel22, Grayling24} for details) is fixed to zero during this first fitting round.
    \item Following the first fitting round, any requested phase cut is reapplied and individual photometric data points which are significant outliers from the best-fit model are optionally cut.
    \item A configurable number of other fitting rounds take place. In the case of SALT, the model-error covariance term is fixed based on the best-fit model parameters from the previous round rather than being recomputed. The result from the final fitting round is returned as the answer, with uncertainties computed based on the final converged optimum solution.
\end{enumerate}

One consequence of the vectorised approach to light curve fitting in \codename{} is that each event must receive the same computations; as SNe are processed in parallel, one cannot receive more optimiser iterations than another, for example. This differs to \snana{}, where serial computation means that each event can be continuously iterated until the optimiser converges. Throughout this work we employ the multi-stage approach to fitting described above for three fitting rounds, running the optimiser for 25 iterations in the first and 15 in the second and third. This is user-customisable, but we have found that the vast majority of light curve fits converge within this optimisation budget and those which do not tend to be cases of poor quality data where parameters are at the prior bound or do not reach a stationary point even with further iterations. As discussed in Section~\ref{sec:val_specz}, our light curve fits show good consistency with \snana{} which serves to justify this choice.

\codename{} also matches \snana{} in supporting light curve fits without a spec-$z$. When fitting with a spec-$z$, redshift uncertainty is neglected and the rest-frame filter response of each SN can be pre-computed based on its redshift. Without a spec-$z$, the redshift is instead inferred jointly with other SN parameters when fitting a light curve. This creates additional computation, as the rest-frame filter response can no longer be pre-computed and must instead be calculated live within the fitting code for every proposed redshift. \codename{} generically supports light curve fitting without a spec-$z$ for any implemented SED model, either with a photo-$z$ prior from host galaxy photometry or with no available redshift information.

\subsubsection{Fitting \textsc{redback-jax} models}
\label{sec:arnett_fitting}

\codename{} also provides support for light curve fitting using physically-motivated models using \textsc{redback-jax} (\textcolor{blue}{Sarin et al., in prep.}). This enables scalable, GPU-accelerated light curve fits with e.g. magnetar models or the Arnett model, though it should be noted that at present this uses the same optimiser as employed in SALT/BayeSN fits rather than the sampling-based approaches implemented in \textsc{redback} \citep{Sarin24} using bilby \citep{Ashton19} and \textsc{redback-jax} using BlackJAX \citep{Cabezas24}. Light curve fitting with \textsc{redback-jax} models is altogether more experimental than SALT/BayeSN fits. 

To demonstrate this, we simulate some light curves and then fit them using a variety of \textsc{redback-jax} models; the Arnett model \citep{Arnett82, Pinto00, Sarin26b}, a magnetar model \citep{Nicholl17, Sarin22} and a combined magnetar plus nickel decay model \citep{Guillochon18}. For fits using the Arnett model, we use flat priors on $M_{\rm ej}$, $f_{\rm Ni}$, $v_{\rm ej}$ and the optical opacity $\kappa$ within the model bounds. The gamma-ray opacity is fixed at 0.03\,cm$^2$\,g$^{-1}$, as in \citet{Sarin26}. We also fix the temperature floor at 4,000\,K, as it only affects later epochs after the photosphere has cooled and is unconstrained for many light curves. For the magnetar models, we use a Gaussian opacity prior, $\kappa=0.10\pm0.02$\,cm$^2$\,g$^{-1}$, and allow the gamma-ray opacity to vary, fixing the temperature floors at 5,000 and 4,500\,K respectively. 

\begin{figure}
    \includegraphics[width=\columnwidth]{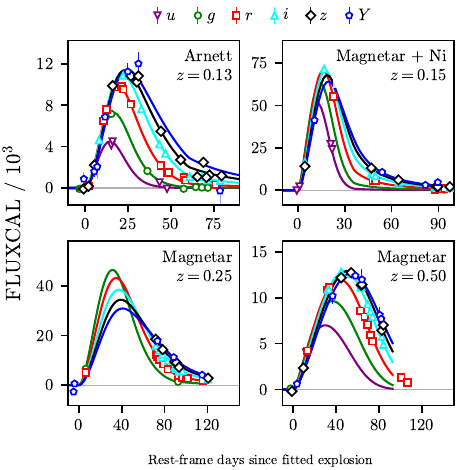}
    \caption{Example Arnett, magnetar and magnetar plus nickel model fits to simulated LSST light curves using the same models.}
    \label{fig:arnett_lightcurves}
\end{figure}

Fig. \ref{fig:arnett_lightcurves} shows some examples of light curves which have been simulated with LSST cadence and then fit using these models, and Table \ref{tab:arnett_recovery} presents the recovery of the true simulated parameters for each of these models. On average parameters are recovered relatively well, with median residuals close to zero and negligible compared to typical uncertainties. We consider the robustness of the uncertainties by assessing the $\sigma_{\rm NMAD}$ of the distribution of residuals normalised by uncertainties, finding that overall these are close to one and uncertainties are relatively well calibrated. However, we also consider the fraction of events where the true value is contained within the 95 per cent confidence interval and find that in some cases coverage is less than 95 per cent, suggesting that there are more outliers than expected. Overall, \codename{} can be used to perform large numbers of these fits at scale, though sampling-based methods can be used for more robust exploration of degeneracies between different parameters.

\begin{table*}
    \caption{Median residual (fitted value - true value), $\sigma_{\rm NMAD}$ of the residuals, $\sigma_{\rm NMAD}$ of the pull distribution and 95 per cent coverage for model fits. \newline
    $^*$$M_{\rm Ni}$ is derived from $M_{\rm ej}$ and $f_{\rm Ni}$, rather than a parameter included in fitting.}
    \label{tab:arnett_recovery}
    \centering
    \small
    \setlength{\tabcolsep}{8pt}
    \begin{tabular}{lrrrr}
        \hline
        Parameter & Median residual & $\sigma_{\rm NMAD}$ & Pull $\sigma_{\rm NMAD}$ & 95\% coverage (\%) \\
        \hline
\textit{Arnett} & & & & \\
$M_{\rm ej}$ ($M_\odot$) & $+7.0\times10^{-4}$ & $0.20$ & 0.86 & 92.1 \\
$f_{\rm Ni}$ & $-7.3\times10^{-5}$ & $0.0088$ & 0.87 & 92.5 \\
$M_{\rm Ni}$ ($M_\odot$)$^*$ & $-2.7\times10^{-5}$ & $0.0030$ & 0.84 & 93.6 \\
$v_{\rm ej}$ ($\mathrm{km\,s^{-1}}$) & $+2.5$ & $140$ & 0.96 & 93.2 \\
$\kappa$ ($\mathrm{cm^2\,g^{-1}}$) & $-1.3\times10^{-4}$ & $0.011$ & 0.88 & 92.5 \\
$t_0$ (d) & $+0.0076$ & $0.21$ & 1.02 & 92.0 \\
\hline
\textit{Magnetar} & & & & \\
$P_0$ (ms) & $-0.018$ & $0.20$ & 0.95 & 88.9 \\
$B_{\rm p}$ ($10^{14}$ G) & $+0.0037$ & $0.14$ & 1.10 & 86.8 \\
$M_{\rm ej}$ ($M_\odot$) & $+0.028$ & $1.4$ & 1.04 & 87.3 \\
$v_{\rm ej}$ ($\mathrm{km\,s^{-1}}$) & $+3.6$ & $99$ & 0.99 & 92.1 \\
$\kappa$ ($\mathrm{cm^2\,g^{-1}}$) & $+1.6\times10^{-4}$ & $0.020$ & 1.05 & 91.7 \\
$\kappa_\gamma$ ($\mathrm{cm^2\,g^{-1}}$) & $+3.1\times10^{-4}$ & $0.0037$ & 0.95 & 88.8 \\
$t_0$ (d) & $+0.019$ & $0.39$ & 1.04 & 89.5 \\
\hline
\textit{Magnetar + nickel} & & & & \\
$P_0$ (ms) & $-0.025$ & $0.39$ & 0.94 & 90.5 \\
$B_{\rm p}$ ($10^{14}$ G) & $+0.0034$ & $0.42$ & 0.91 & 92.8 \\
$M_{\rm ej}$ ($M_\odot$) & $+0.032$ & $0.53$ & 1.05 & 85.0 \\
$f_{\rm Ni}$ & $-0.0040$ & $0.047$ & 1.05 & 85.6 \\
$v_{\rm ej}$ ($\mathrm{km\,s^{-1}}$) & $+0.025$ & $74$ & 0.96 & 93.8 \\
$\kappa$ ($\mathrm{cm^2\,g^{-1}}$) & $-9.8\times10^{-4}$ & $0.019$ & 1.13 & 86.7 \\
$\kappa_\gamma$ ($\mathrm{cm^2\,g^{-1}}$) & $-3.5\times10^{-5}$ & $0.0079$ & 1.00 & 86.1 \\
$t_0$ (d) & $+0.0010$ & $0.073$ & 1.00 & 92.5 \\
\hline

    \end{tabular}
\par\vspace{3pt}
\end{table*}

\subsection{Photometric classification}

\codename{} provides support for the use of external photometric classification packages as part of an analysis pipeline in a modular way, allowing users to build in support for new classifiers. At the present time support for \textsc{supernnova} \citep{Moller20} is built in. \codename{} allows for samples of simulated events to be directly used to train photometric classification models or load existing trained models and apply them for classification. This is a vital step for cosmological analyses with photometrically-classified samples \citep[e.g.][]{Hlozek12, Jones17, Jones18b, Chen22, DES24, Popovic24Amalgame, Vincenzi24, Popovic25}.

\subsection{Distance fitting and cosmological inference}
\label{sec:bbc}

\begin{figure}
	\includegraphics[width=\columnwidth]{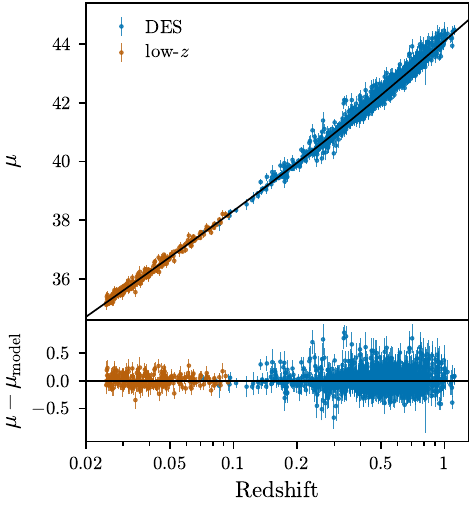}
    \caption{An example bias-corrected Hubble diagram for a simulated DES-SN5YR sample obtained from applying light curve fitting and \textsc{salt2mu} using \codename{}.}
    \label{fig:hubble_diagram}
\end{figure}

\begin{figure}
	\includegraphics[width=\columnwidth]{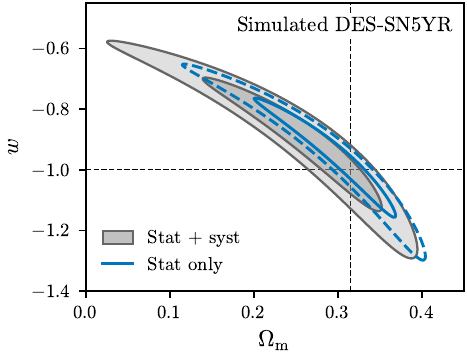}
    \caption{Inferred cosmological constraints from the simulated Hubble diagram presented in Fig.~\ref{fig:hubble_diagram}.}
    \label{fig:cosmology_contours}
\end{figure}

Beyond simulations and light curve fitting, \codename{} includes a \jax{} implementation of the \textsc{salt2mu} and \textsc{wfit} codes in \snana{}. For many SN~Ia models, standardisation to estimate distance is a distinct step to light curve fitting. BayeSN is a notable exception to this, as it jointly infers distance alongside other SN parameters. For typical SALT analyses \citep[e.g.][]{Brout22, DES24, Popovic25, Rigault25}, SNe are standardised according to the Tripp formula \citep{Tripp98}:

\begin{equation}
    \mu = m_B - M + \alpha x_1 - \beta c + \gamma_{\rm host} - \Delta\mu_{\rm bias},
\end{equation}
where $m_B$, $x_1$, and $c$ are the peak $B$-band apparent magnitude, stretch and colour parameters from light curve fitting, $M$ is a population-level absolute magnitude and $\alpha$ and $\beta$ are linear standardisation coefficients. $\gamma_{\rm host}$ is a host-galaxy standardisation term which typically corrects for the `mass-step' \citep{Kelly10, Sullivan10, Lampeitl10}. $\Delta\mu_{\rm bias}$ is a simulation-derived bias correction which is sometimes used, though some analyses opt to include selection effects within hierarchical Bayesian models \citep[e.g. UNITY;][]{Rubin15, Rubin25} and \citet{Rigault25, Ginolin25} consider a volume-limited sample of SNe and hence omit a selection effect correction.

\textsc{salt2mu} is an implementation of this standardisation procedure in \snana{}, using the approach for standardisation outlined in \citet{Marriner11}, an approach designed to ensure the standardisation parameters are insensitive to the cosmology assumed during this process\footnote{This is distinct from the approach used in \citet{Boyd26}, where these are inferred jointly with cosmological parameters.}. This approach is combined with the BBC framework outlined in \citet{Kessler17} to estimate distances from fitted light curve parameters, corrected for population-level selection bias and marginalised over the impact of photometric contamination\footnote{This does mean that these distances should not be analysed in sub-samples of the population, as discussed in \citet{Vincenzi25}.}.

In previous cosmological analyses, \snana{} is used to estimate distances which are then used in an external cosmological inference framework; DES-SN5YR used the \textsc{cosmosis} framework \citep{DES24, Zuntz15}. However \snana{} does provide a built-in \textsc{wfit} module which can be used to obtain quick cosmological constraints for testing using a grid search. \codename{} implements this and extends this, providing support for cosmological parameter inference using either a grid search approach, a Laplace approximation using an optimiser or a full MCMC approach using the No U-Turn Sampler \citep[NUTS;][]{Hoffman11} implemented in \textsc{numpyro} \citep{Phan19}.

Fig. \ref{fig:hubble_diagram} presents an example of the \codename{} implementation of \textsc{salt2mu}, showing a Hubble diagram obtained from a simulated DES-SN5YR sample after standardisation. Meanwhile, Fig. \ref{fig:cosmology_contours} presents cosmological contours obtained from applying the \jax{} \textsc{wfit} implementation to this simulated dataset.

\subsection{Modular analysis pipelines}

\codename{} has been designed to allow the user to develop custom end-to-end analysis pipelines, following a simple modular design. This allows the various supported features - light curve simulations, fitting, photometric classification, standardisation and cosmological inference - all to run inside one single \textsc{python} script. This supports multiple different datasets within the same script, for example combining separate high- and low-redshift surveys, alongside support for incorporating different systematic tests within the pipeline to construct a systematic uncertainty budget. This is analogous to the \textsc{pippin} \textsc{python} wrapper \citep{Hinton20} often used for \snana{} analyses, which defines the full cosmological inference pipeline in a single input file and will deploy jobs on HPC, merge together the outputs and pass them along to the next stage. The advantage of running all stages in one single script is that all stages can share compiled functions so that only one batch will be compiled for light curve simulation, fitting and standardisation across the full analysis, maximising performance. For compatibility with \snana{}, \codename{} is able to construct \textsc{python} pipelines directly from \textsc{pippin} input files. In future, it may also prove desirable to allow \textsc{pippin} to interact with \codename{} to deploy jobs on HPC, though the high performance of \codename{} makes running a full end-to-end pipeline at large scale possible within a single process in only a few hours of GPU time.

\subsection{Interfacing with SBI applications}

One of the main motivations of \codename{} is to support the continued development and application of SBI methods for analysis of populations of transient events. As demonstrated in Section~\ref{sec:performance}, \codename{} provides a fully JIT-compiled simulation pipeline with very fast performance. A key advantage of this code is that the \jax{} implementation of the simulator can directly interface with various SBI tools, including \textsc{sbijax} \citep{Dirmeier24}, \textsc{sbiax} \citep{Homer25} and \textsc{flowjax} \citep{FlowJAX}. \codename{} has a dedicated method for constructing an SBI pipeline for a given SN model, allowing full end-to-end simulations including host galaxy association and application of selection effects to be directly included within the training loop for an SBI algorithm, creating on-the-fly simulations.

\subsection{Differentiability}

As previously mentioned, \codename{} supports autodiff through \jax{} computations, valuable for tasks such as light curve fitting given this enables gradient-based parameter exploration. This, however, does not make full use of the potential benefits of autodiff. In principle, a full \jax{} simulation pipeline could enable differentiable simulations, with a number of key advantages. For example, SBI methods can use the gradients of the simulator to learn the posterior from far fewer simulations \citep{Brehmer20, Zeghal22}. However, within this work, we focus primarily on applying \codename{} to the existing supernova cosmology pipeline as used for \citet{DES24, Vincenzi24, Popovic25, Camilleri26, Lee26}. Much of the light curve simulation pipeline is differentiable; gradients can be passed through SED generation, photometry and noise. However, the pipeline does involve a number of discrete steps, for example applying selection effects, host association and observation scheduling. Moving towards a fully differentiable simulator is beyond the scope of this work but is planned for the future.

\section{Validation against \snana{}}
\label{sec:validation}

It is of course essential that \codename{} is able to provide consistent results with \snana{} with the same inputs. Within this section we focus on validation of the results obtained by \codename{}. 

For simulations, although \codename{} has been designed to match \snana{}\footnote{In rare cases \codename{} deliberately diverges from \snana{} to use more accurate computational methods; these are documented at \url{https://jax-snana.readthedocs.io}}, it is not possible to achieve exact parity as the random number generation sequence does not match between the two. Instead, parity between the simulations is assessed in two key ways. The first is looking for statistical consistency between large scale inputs across a wide variety of model parameters. The second uses a bespoke method which uses the model parameters from an \snana{} simulation directly as input to \codename{}, allowing for testing to ensure that both give consistent outputs given identical model parameters.

For light curve fitting, exact parity between the two codes is not the ambition; \snana{} uses the \textsc{minuit} optimiser \citep{James75} while \codename{} uses the custom Levenberg-Marquardt optimiser outlined in Section \ref{sec:fit}. In general we would expect very similar results but over large samples of noisy data different optimisers are inevitably prone to getting stuck in different local minima on occasion. We compare the outputs of both codes with the same inputs and assess both consistency between the two and recovery of true simulated parameters.

\subsection{Simulation parity}
\label{sec:val_lsst}

We begin by comparing simulations of supernova populations in \snana{} and \codename{} with identical input parameters. We consider this in two ways; we expect noiseless model photometry to match effectively perfectly, with minor differences due to different numeric integration methods between the codes, and we expect statistical consistency in the populations which pass selection effects. There is an element of randomness in which SNe are selected by the detection efficiency pipeline, as discussed in Section \ref{sec:searcheff}, but we expect similar results.

Table~\ref{tab:replay} shows the mean and standard deviation of the photometric residuals between \codename{} and \snana{} for the same SED models with identical SN parameters and epochs, quoted in mmag. Overall agreement is very good between the codes, with the mean residual negligible at less than 1 mmag across the various models. For SN~Ia models, there is also a small scatter in the residual across all photometric epochs, typically less than or close to 1 mmag. The scatter is slightly increased in some of the other models to a few mmag, though this is unsurprising as there is a slight methodological difference in how \codename{} treats these templates compared to \snana{}. \snana{} pre-integrates the filters at different redshifts to construct a grid and interpolate the grid for the simulated phases and redshifts to get model photometry. \codename{} instead evaluates the interpolated model SED at the simulated phases and uses the exact redshift, without pre-integrating the grid; this slight difference in approach makes this higher scatter unsurprising. \codename{} uses this different approach for increased accuracy and for consistency with how the SN~Ia models are treated within the code, and still achieves very high performance when simulating from these templates. Overall, \codename{} is shown to have a high level of numerical accuracy and consistency with \snana{}.

\begin{figure*}
	\includegraphics[width=\textwidth]{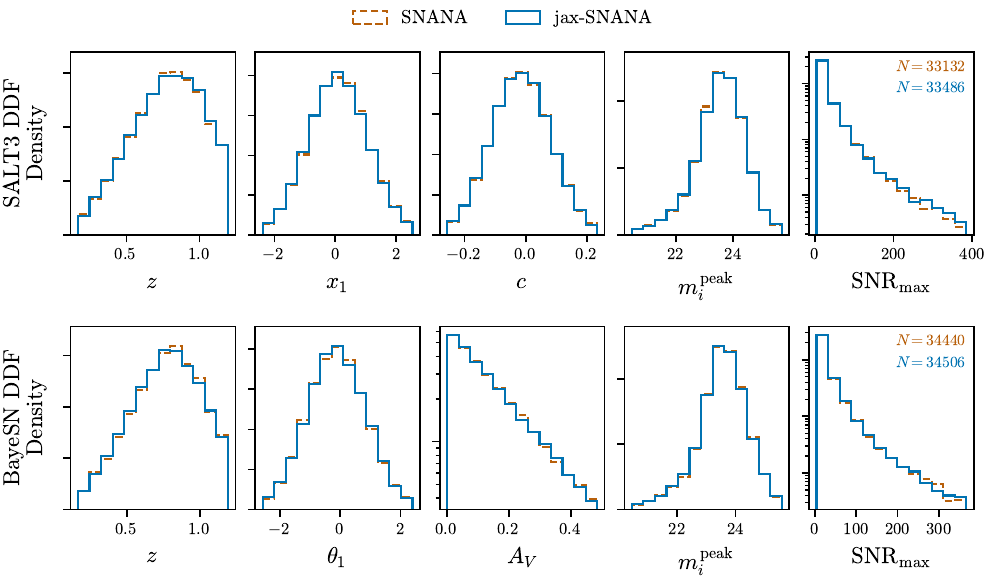}
    \caption{Distributions of simulated SN parameters compared between \codename{} and \snana{}, for LSST simulations with the same inputs, after applying selection effects. While these are not identical due to random number generation differences, overall these are very consistent. The top panel shows a SALT simulation, while the bottom panel shows a BayeSN simulation.}
    \label{fig:sim_population_lsst}
\end{figure*}

\begin{figure*}
	\includegraphics[width=\textwidth]{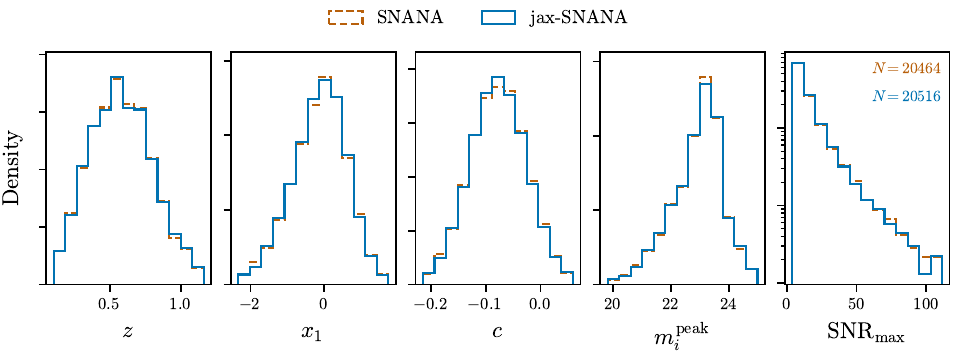}
    \caption{As Fig.~\ref{fig:sim_population_lsst} but showing distributions of parameter values after selection effects for simulations using the same configuration as the DES-SN5YR analysis. Again, we see excellent consistency between \snana{} and \codename{}.}
    \label{fig:des5yr_population}
\end{figure*}

We next compare the outputs of different simulations using each code with the same input files, rather than with identical parameters as in the previous case. This is a full consistency test of both codes through parameter sampling, simulation and selection. Fig. \ref{fig:sim_population_lsst} presents distributions of simulated parameter values and observational properties after applying selection effects to \snana{} and \codename{} simulations of 100,000 SNe~Ia in the LSST DDF with the same input files; examples are shown for both SALT and BayeSN. These demonstrate the excellent consistency we see between the two. This first example used relatively simple simulations without any treatment of host-dependence, but Fig. \ref{fig:des5yr_population} presents distributions obtained from simulated samples using the same simulation inputs as for the DES-SN5YR analysis \citep{Kessler19DES, Vincenzi24} including the link between properties of SNe~Ia and their host galaxies \citep{Popovic23}, showing excellent consistency with more realistic and complex simulations. Table~\ref{tab:cutflow} presents the number of SNe rejected by various selection cuts and the total accepted for a variety of \codename{} and \snana{} simulations, further reiterating the consistency between the two.

\begin{table*}
	\centering
	\caption{Mean and standard deviation of the residual differences in photometry simulated in \snana{} and \codename{}, comparing matched epochs with the same observing schedules and simulated parameters. The values are quoted in mmag, and demonstrate very precise photometric agreement between the two codes.}
	\label{tab:replay}
	\setlength{\tabcolsep}{5pt}
	\begin{tabular}{lcccccc}
		\hline
		& \multicolumn{6}{c}{Mean (std) $m_{\text{\codename{}}} - m_{\rm SNANA}$ / mmag}\\
		Sample & $u$ & $g$ & $r$ & $i$ & $z$ & $Y$\\
		\hline
		LSST DDF, \salt{}3 & \saltReplayDdfCellU & \saltReplayDdfCellG & \saltReplayDdfCellR & \saltReplayDdfCellI & \saltReplayDdfCellZ & \saltReplayDdfCellY\\
		LSST WFD, \salt{}3 & \saltReplayWfdCellU & \saltReplayWfdCellG & \saltReplayWfdCellR & \saltReplayWfdCellI & \saltReplayWfdCellZ & \saltReplayWfdCellY\\
		LSST DDF, \bayesn{} & \bayesnReplayDdfCellU & \bayesnReplayDdfCellG & \bayesnReplayDdfCellR & \bayesnReplayDdfCellI & \bayesnReplayDdfCellZ & \bayesnReplayDdfCellY\\
		LSST WFD, \bayesn{} & \bayesnReplayWfdCellU & \bayesnReplayWfdCellG & \bayesnReplayWfdCellR & \bayesnReplayWfdCellI & \bayesnReplayWfdCellZ & \bayesnReplayWfdCellY\\
		LSST DDF, SN II & \contamReplaySniiCellU & \contamReplaySniiCellG & \contamReplaySniiCellR & \contamReplaySniiCellI & \contamReplaySniiCellZ & \contamReplaySniiCellY\\
		LSST DDF, SN 1991bg-like & \contamReplayNinetyOneBgCellU & \contamReplayNinetyOneBgCellG & \contamReplayNinetyOneBgCellR & \contamReplayNinetyOneBgCellI & \contamReplayNinetyOneBgCellZ & \contamReplayNinetyOneBgCellY\\
		LSST DDF, SN Iax & \contamReplayIaxCellU & \contamReplayIaxCellG & \contamReplayIaxCellR & \contamReplayIaxCellI & \contamReplayIaxCellZ & \contamReplayIaxCellY\\
		DES-SN5YR, \salt{}3 & -- & \desReplayCellG & \desReplayCellR & \desReplayCellI & \desReplayCellZ & --\\
		\hline
	\end{tabular}
\end{table*}

\begin{table*}
	\centering
	\caption{Numbers of SNe rejected by various selection cuts, and the number accepted, for large simulated samples in \codename{} and \snana{} with identical inputs, demonstrating statistical consistency between the two. Due to differences in random number generation between the two, these will not agree exactly but should be consistent for large samples. Numbers are separately quoted for \snana{}/\codename{}.}
	\label{tab:cutflow}
	\setlength{\tabcolsep}{4pt}
	\begin{tabular}{lccccc}
		\hline
		Sample & $N_{\rm gen}$ & $N_{\rm epoch}$ & Detection & Cuts & Accepted\\
		\hline
		LSST DDF, \salt{}3 & \saltPopDdfNGen & \saltPopDdfRejNepSnana\,/\,\saltPopDdfRejNepJax & \saltPopDdfRejSearcheffSnana\,/\,\saltPopDdfRejSearcheffJax & \saltPopDdfRejCutwinSnana\,/\,\saltPopDdfRejCutwinJax & \saltPopDdfNAccSnana\,/\,\saltPopDdfNAccJax\\
		LSST WFD, \salt{}3 & \saltPopWfdNGen & \saltPopWfdRejNepSnana\,/\,\saltPopWfdRejNepJax & \saltPopWfdRejSearcheffSnana\,/\,\saltPopWfdRejSearcheffJax & \saltPopWfdRejCutwinSnana\,/\,\saltPopWfdRejCutwinJax & \saltPopWfdNAccSnana\,/\,\saltPopWfdNAccJax\\
		LSST DDF, \bayesn{} & \bayesnPopDdfNGen & \bayesnPopDdfRejNepSnana\,/\,\bayesnPopDdfRejNepJax & \bayesnPopDdfRejSearcheffSnana\,/\,\bayesnPopDdfRejSearcheffJax & \bayesnPopDdfRejCutwinSnana\,/\,\bayesnPopDdfRejCutwinJax & \bayesnPopDdfNAccSnana\,/\,\bayesnPopDdfNAccJax\\
		LSST WFD, \bayesn{} & \bayesnPopWfdNGen & \bayesnPopWfdRejNepSnana\,/\,\bayesnPopWfdRejNepJax & \bayesnPopWfdRejSearcheffSnana\,/\,\bayesnPopWfdRejSearcheffJax & \bayesnPopWfdRejCutwinSnana\,/\,\bayesnPopWfdRejCutwinJax & \bayesnPopWfdNAccSnana\,/\,\bayesnPopWfdNAccJax\\
		DES-SN5YR, \salt{}3 & \desPopNGen & \desPopRejNepSnana\,/\,\desPopRejNepJax & \desPopRejSearcheffSnana\,/\,\desPopRejSearcheffJax & \desPopRejCutwinSnana\,/\,\desPopRejCutwinJax & \desPopNAccSnana\,/\,\desPopNAccJax\\
		\hline
	\end{tabular}
\end{table*}

\subsection{Light curve fitting}
\label{sec:val_specz}

We now consider light curve fitting parity between \snana{} and \codename{}. This is focused purely on SALT fits; although both codes include an implementation of BayeSN for light curve fitting, \snana{}'s had slightly reduced functionality compared to the full BayeSN model as it did not incorporate the intrinsic scatter distribution ($\Sigma_\epsilon$ in \citealt{Mandel22}). \codename{} does include this term when fitting light curves with the BayeSN model, so an exact parity test is not possible. Additionally, \codename{} focuses on optimisation, rather than the sampling or variational-inference approaches used in the public BayeSN code \citep{Grayling24, Uzsoy24} which are likely to provide more robust posterior estimation and uncertainty.

\subsubsection{Fitting with spectroscopic redshift}

\begin{figure*}
	\includegraphics[width=\textwidth]{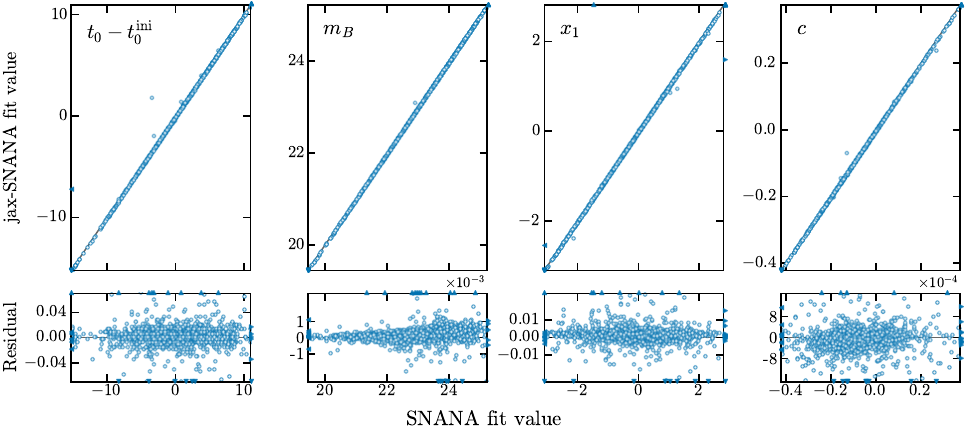}
    \caption{\textbf{Top}: Comparison of fitted values for SALT fits to the same simulated SNe using both \codename{} and \snana{} to establish parity between the two, using spec-$z$. \textbf{Bottom}: Residuals of the top panels.}
    \label{fig:fit_parity_specz}
\end{figure*}

\begin{figure*}
	\includegraphics[width=\textwidth]{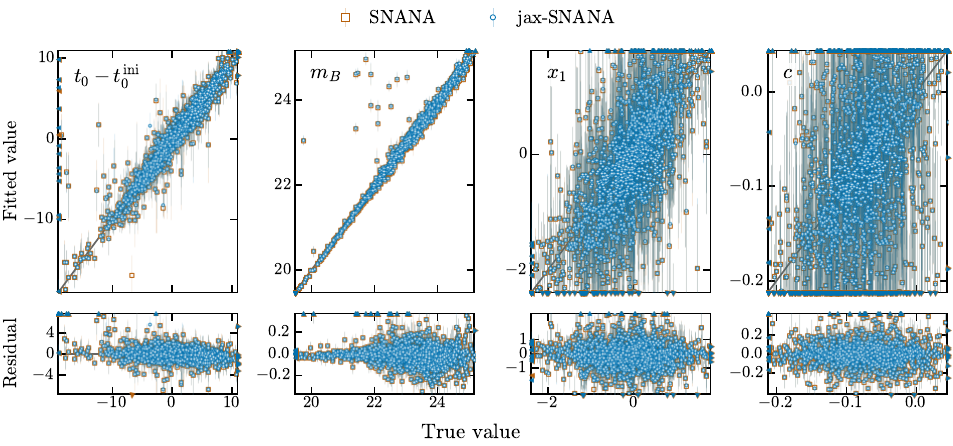}
    \caption{\textbf{Top}: Comparison between fitted values and truth for SALT fits, to the same simulated SNe presented in Fig.~\ref{fig:fit_parity_specz}, using both \codename{} and \snana{}. \textbf{Bottom}: Residuals of the top panels.}
    \label{fig:fit_recovery_specz}
\end{figure*}

We first consider fits which assume spec-$z$s are available for the SN, comparing the best-fit solutions between the two codes. We consider both fit parity and recovery of true simulated parameter values. Figure \ref{fig:fit_parity_specz} presents the parameter values inferred from fitting the same SALT simulation of a sample of \fitSpeczNSalt{} DES-SN5YR SN~Ia light curves with \codename{} and SNANA, while Fig.~\ref{fig:fit_recovery_specz} presents the residuals compared to truth for both codes. Table \ref{tab:fitparity} presents the median residual between the two along with the associated scatter. We consider fits of peak B-band apparent magnitude, $m_B$, along with $x_1$, $c$ and time-of-maximum $t_0$, which is plotted as a shift relative to the fiducial value estimated purely from the data (the initial estimate of the time of maximum, from the observations around the brightest measured flux). Overall there is very close agreement between the two codes, with just \fitSpeczSaltNOutlier{} outliers where the two find different local minima and consistent scatter around the truth from both. Overall, \codename{} is shown to produce robust light curve fits consistent with SNANA.

Table \ref{tab:fitparity} also presents uncertainty metrics from both fitters; the median ratio of the uncertainties, and the $\sigma_{\rm NMAD}$ of the pull distribution to assess whether the uncertainties are well calibrated. Overall, $m_B$ and $x_1$ have well-calibrated uncertainties with $\sigma_{\rm NMAD}$ very close to 1 for both codes. Both are also consistent for $t_0$ and $c$ in pull, although both have overestimated uncertainties for $t_0$ and underestimated for $c$. Uncertainties from \codename{} are based on the Hessian evaluated at the optimiser solution using autodiff. \snana{} features an approximate Hessian-based uncertainty, and optionally a more robust uncertainty obtained from a profile of the likelihood around the fitted solution. Table \ref{tab:fitparity} presents \snana{} results obtained using the profile uncertainty, which we find gives very consistent results with the exact Hessian uncertainty in \codename{}; this adds additional performance overhead in \snana{} beyond the Hessian approach. The lack of calibrated uncertainties for $t_0$ is driven by a mismatch between the model used for simulating and fitting; when fitting, \snana{} and \codename{} include an error term associated with scatter in the SALT model training itself, whereas the simulations use the SALT model directly with no additional scatter. While we could disable these error terms to get a pull $\sigma_{\rm NMAD}$ closer to 1, this would not be reflective of what is used for real analyses. The underestimated uncertainty for $c$ is seen across both codes and likely arises from the optimiser-based approach being used; the key result is the consistency between \snana{} and \codename{}.

\subsubsection{Fitting with photometric redshifts}
\label{sec:val_photoz}

We next consider fits for SNe which do not have spec-$z$s available. We consider three variants to consider cases where different redshift information is available:

\begin{itemize}
    \item We first consider a case where no redshift information is available a priori for the SNe, meaning an uninformative uniform prior is used in the light curve fit.
    \item We next consider a case where each SN has a Gaussian photo-$z$ prior, a slightly simplistic treatment for SNe where some redshift information is available from its host galaxy photometry. The prior is centred on the host photo-$z$ point estimate with width set by its quoted uncertainty, as in \citet{Mitra23} and the Gaussian-prior case of \citet{ChenR24}; here the point estimate is the median of the host photo-$z$ PDF and the width is half its 16--84 per cent range.
    \item We finally consider a case where each SN has a photo-$z$ PDF from host galaxy photometry expressed using quantiles \citep{Malz18}, the most realistic test. We use the ELAsTiCC SN~Ia host-galaxy library \citep{ELAsTiCC23, Mitra25}, built with the GHOST host-association framework \citep{Gagliano21}. Each galaxy in the catalogue carries a photo-$z$ PDF from a normalising flow trained with PZFlow \citep{Crenshaw24} and stored as 11 evenly-spaced quantiles of the cumulative distribution. 
\end{itemize}

\begin{figure*}
	\includegraphics[width=\textwidth]{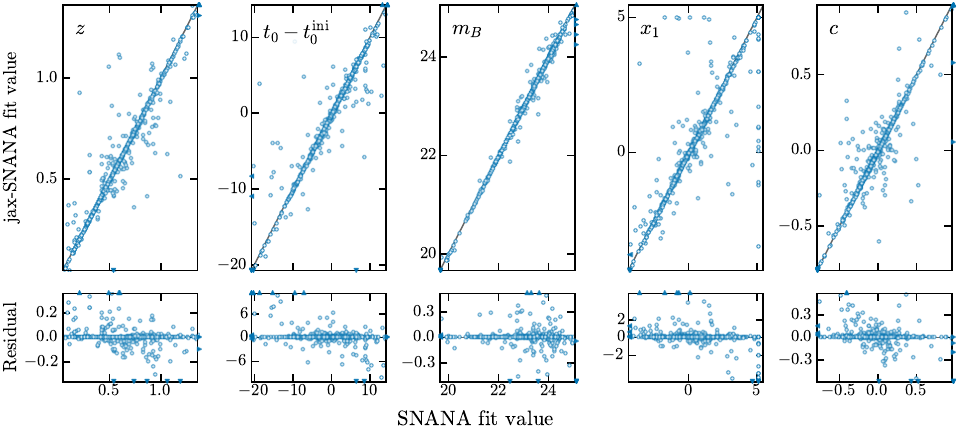}
    \caption{\textbf{Top}: Comparison of fitted values for SALT fits, with quantile photo-$z$ priors and redshift a fitted parameter alongside SALT parameters, to the same simulated SNe using both \codename{} and \snana{} to establish parity between the two. \textbf{Bottom}: Residuals of the top panels.}
    \label{fig:fit_parity_photoz}
\end{figure*}

\begin{figure*}
	\includegraphics[width=\textwidth]{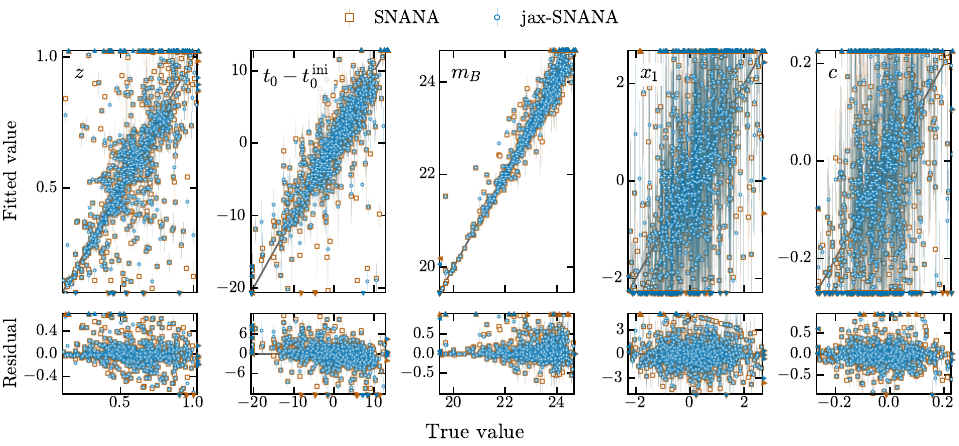}
    \caption{\textbf{Top}: Comparison between fitted values and truth for SALT fits with quantile photo-$z$ priors, to the same simulated SNe presented in Fig.~\ref{fig:fit_parity_photoz}, using both \codename{} and \snana{}. \textbf{Bottom}: Residuals of the top panels.}
    \label{fig:fit_recovery_photoz}
\end{figure*}

Fig. \ref{fig:fit_parity_photoz} shows the inferred parameter values from fitting the same simulated light curves with both \snana{} and \codename{}, with a quantile photo-$z$ prior, while Fig.~\ref{fig:fit_recovery_photoz} presents residuals from the truth for both codes; overall, these demonstrate an excellent level of consistency between the two. Similar plots for the other two photometric cases are presented in Appendix~\ref{app:photoz_priors}. Table \ref{tab:fitparity} presents a variety of metrics describing consistency between the values and uncertainties from each code, including the median residual between fits using the two codes and the scatter. 

Considering the fitted parameter values, there is an overall good level of consistency between the two, although photo-$z$ fits are somewhat more prone to getting trapped in local minima, with more outliers between the two. Some of these are caused by the two fitting different photometric data; \textsc{minuit} will sometimes move away from the true $t_0$ during the first fitting round, hence many data points are removed by the phase cut after the first round. \codename{} therefore is used to fit different photometric epochs to \snana{} even though both are applied to the same events. These are typically poor fits to low quality data when neither code recovers the true simulated value.

The greater challenge with photo-$z$ fitting is that the parameter space is more complex than for spec-$z$ fitting, with strong degeneracies between parameters such as $c$ and redshift. This creates an additional challenge when estimating uncertainties; the optimiser-based approaches used by \codename{} and \snana{} lead to poorly-calibrated uncertainties in photo-$z$ fitting, as shown in Table \ref{tab:fitparity}. In all cases the $\sigma_{\rm NMAD}$ of the pull distribution is larger than 1, most notably for the inferred redshift. This demonstrates that the uncertainties from these codes are underestimated, and suggests that sampling-based approaches should be used to properly account for the complex degeneracies in the parameter space. Nevertheless, our objective in this work is parity with \snana{}, and hence we focus on that. In Section~\ref{sec:val_specz}, when exploring fitting with a spec-$z$, we found good agreement between our Hessian-based uncertainties and \snana{}'s more complex profile-based uncertainties. This was not the case, however, for photo-$z$ fitting, most notably for the redshift parameter itself. \codename{} does support a similar profile-based uncertainty across all parameters as \snana{} does but at increased computational expense. As a compromise, we have implemented an option to profile the likelihood and obtain a more robust uncertainty only for redshift, with a far smaller impact on performance; the results presented in Table \ref{tab:fitparity} use this approach. We still see a good level of consistency between the uncertainties from the two codes, with the median ratio of the uncertainties from both close to 1. Considering both fitted parameters and their uncertainties, \codename{} and \snana{} are in close agreement for light curve fitting with photo-$z$s.

\begin{table*}
	\centering
	\caption{Light curve fit parity between \codename{} and \snana{} for spec-$z$s and each photo-$z$ prior. The first two columns give the median and $\sigma_{\rm NMAD}$ of the residual between the two codes as a fraction of the \snana{} uncertainty. The third presents the median ratio of the \codename{} and \snana{} uncertainties, while the fourth and fifth present $\sigma_{\rm NMAD}$ of the fitted values around the truth for both codes. Finally, the last two columns present the $\sigma_{\rm NMAD}$ of the pull ($({\rm fit} - {\rm true})/\sigma$) from both codes; for perfectly-calibrated uncertainties, these should be exactly 1.}
	\label{tab:fitparity}
	\begin{tabular}{lccccccc}
		\hline
		& \multicolumn{2}{c}{$\Delta / \sigma_{\rm SNANA}$} & & \multicolumn{2}{c}{$\sigma_{\rm NMAD}({\rm fit} - {\rm true})$} & \multicolumn{2}{c}{$\sigma_{\rm NMAD}[({\rm fit} - {\rm true})/\sigma]$}\\
		Parameter & Median & $\sigma_{\rm NMAD}$ & $\sigma_{\text{\codename{}}}/\sigma_{\rm SNANA}$ & \snana{} & \codename{} & \snana{} & \codename{}\\
		\hline
		\multicolumn{8}{l}{Spectroscopic $z$: $N = \fitSpeczNSalt$}\\
		$t_0$ (d) & $\fitSpeczSaltTzeroPullMedSci$ & $\fitSpeczSaltTzeroPullSigSci$ & \fitSpeczSaltTzeroSigRatio & ${\fitSpeczSaltTzeroMadSnana}$ & ${\fitSpeczSaltTzeroMadJax}$ & \fitSpeczSaltTzeroTruthPullSnana & \fitSpeczSaltTzeroTruthPullJax\\
		$m_B$ (mag) & $\fitSpeczSaltMBPullMedSci$ & $\fitSpeczSaltMBPullSigSci$ & \fitSpeczSaltMBSigRatio & ${\fitSpeczSaltMBMadSnana}$ & ${\fitSpeczSaltMBMadJax}$ & \fitSpeczSaltMBTruthPullSnana & \fitSpeczSaltMBTruthPullJax\\
		$x_1$ & $\fitSpeczSaltXonePullMedSci$ & $\fitSpeczSaltXonePullSigSci$ & \fitSpeczSaltXoneSigRatio & ${\fitSpeczSaltXoneMadSnana}$ & ${\fitSpeczSaltXoneMadJax}$ & \fitSpeczSaltXoneTruthPullSnana & \fitSpeczSaltXoneTruthPullJax\\
		$c$ & $\fitSpeczSaltCPullMedSci$ & $\fitSpeczSaltCPullSigSci$ & \fitSpeczSaltCSigRatio & ${\fitSpeczSaltCMadSnana}$ & ${\fitSpeczSaltCMadJax}$ & \fitSpeczSaltCTruthPullSnana & \fitSpeczSaltCTruthPullJax\\
		\hline
		\multicolumn{8}{l}{Photometric $z$, no prior: $N = \fitPhotozNopriorN$}\\
		$z$ & $\fitPhotozNopriorZPullMedSci$ & $\fitPhotozNopriorZPullSigSci$ & \fitPhotozNopriorZSigRatio & ${\fitPhotozNopriorZMadSnana}$ & ${\fitPhotozNopriorZMadJax}$ & \fitPhotozNopriorZTruthPullSnana & \fitPhotozNopriorZTruthPullJax\\
		$t_0$ (d) & $\fitPhotozNopriorTzeroPullMedSci$ & $\fitPhotozNopriorTzeroPullSigSci$ & \fitPhotozNopriorTzeroSigRatio & ${\fitPhotozNopriorTzeroMadSnana}$ & ${\fitPhotozNopriorTzeroMadJax}$ & \fitPhotozNopriorTzeroTruthPullSnana & \fitPhotozNopriorTzeroTruthPullJax\\
		$m_B$ (mag) & $\fitPhotozNopriorMBPullMedSci$ & $\fitPhotozNopriorMBPullSigSci$ & \fitPhotozNopriorMBSigRatio & ${\fitPhotozNopriorMBMadSnana}$ & ${\fitPhotozNopriorMBMadJax}$ & \fitPhotozNopriorMBTruthPullSnana & \fitPhotozNopriorMBTruthPullJax\\
		$x_1$ & $\fitPhotozNopriorXonePullMedSci$ & $\fitPhotozNopriorXonePullSigSci$ & \fitPhotozNopriorXoneSigRatio & ${\fitPhotozNopriorXoneMadSnana}$ & ${\fitPhotozNopriorXoneMadJax}$ & \fitPhotozNopriorXoneTruthPullSnana & \fitPhotozNopriorXoneTruthPullJax\\
		$c$ & $\fitPhotozNopriorCPullMedSci$ & $\fitPhotozNopriorCPullSigSci$ & \fitPhotozNopriorCSigRatio & ${\fitPhotozNopriorCMadSnana}$ & ${\fitPhotozNopriorCMadJax}$ & \fitPhotozNopriorCTruthPullSnana & \fitPhotozNopriorCTruthPullJax\\
		\hline
		\multicolumn{8}{l}{Photometric $z$, Gaussian prior: $N = \fitPhotozGaussN$}\\
		$z$ & $\fitPhotozGaussZPullMedSci$ & $\fitPhotozGaussZPullSigSci$ & \fitPhotozGaussZSigRatio & ${\fitPhotozGaussZMadSnana}$ & ${\fitPhotozGaussZMadJax}$ & \fitPhotozGaussZTruthPullSnana & \fitPhotozGaussZTruthPullJax\\
		$t_0$ (d) & $\fitPhotozGaussTzeroPullMedSci$ & $\fitPhotozGaussTzeroPullSigSci$ & \fitPhotozGaussTzeroSigRatio & ${\fitPhotozGaussTzeroMadSnana}$ & ${\fitPhotozGaussTzeroMadJax}$ & \fitPhotozGaussTzeroTruthPullSnana & \fitPhotozGaussTzeroTruthPullJax\\
		$m_B$ (mag) & $\fitPhotozGaussMBPullMedSci$ & $\fitPhotozGaussMBPullSigSci$ & \fitPhotozGaussMBSigRatio & ${\fitPhotozGaussMBMadSnana}$ & ${\fitPhotozGaussMBMadJax}$ & \fitPhotozGaussMBTruthPullSnana & \fitPhotozGaussMBTruthPullJax\\
		$x_1$ & $\fitPhotozGaussXonePullMedSci$ & $\fitPhotozGaussXonePullSigSci$ & \fitPhotozGaussXoneSigRatio & ${\fitPhotozGaussXoneMadSnana}$ & ${\fitPhotozGaussXoneMadJax}$ & \fitPhotozGaussXoneTruthPullSnana & \fitPhotozGaussXoneTruthPullJax\\
		$c$ & $\fitPhotozGaussCPullMedSci$ & $\fitPhotozGaussCPullSigSci$ & \fitPhotozGaussCSigRatio & ${\fitPhotozGaussCMadSnana}$ & ${\fitPhotozGaussCMadJax}$ & \fitPhotozGaussCTruthPullSnana & \fitPhotozGaussCTruthPullJax\\
		\hline
		\multicolumn{8}{l}{Photometric $z$, quantile prior: $N = \fitPhotozQuantN$}\\
		$z$ & $\fitPhotozQuantZPullMedSci$ & $\fitPhotozQuantZPullSigSci$ & \fitPhotozQuantZSigRatio & ${\fitPhotozQuantZMadSnana}$ & ${\fitPhotozQuantZMadJax}$ & \fitPhotozQuantZTruthPullSnana & \fitPhotozQuantZTruthPullJax\\
		$t_0$ (d) & $\fitPhotozQuantTzeroPullMedSci$ & $\fitPhotozQuantTzeroPullSigSci$ & \fitPhotozQuantTzeroSigRatio & ${\fitPhotozQuantTzeroMadSnana}$ & ${\fitPhotozQuantTzeroMadJax}$ & \fitPhotozQuantTzeroTruthPullSnana & \fitPhotozQuantTzeroTruthPullJax\\
		$m_B$ (mag) & $\fitPhotozQuantMBPullMedSci$ & $\fitPhotozQuantMBPullSigSci$ & \fitPhotozQuantMBSigRatio & ${\fitPhotozQuantMBMadSnana}$ & ${\fitPhotozQuantMBMadJax}$ & \fitPhotozQuantMBTruthPullSnana & \fitPhotozQuantMBTruthPullJax\\
		$x_1$ & $\fitPhotozQuantXonePullMedSci$ & $\fitPhotozQuantXonePullSigSci$ & \fitPhotozQuantXoneSigRatio & ${\fitPhotozQuantXoneMadSnana}$ & ${\fitPhotozQuantXoneMadJax}$ & \fitPhotozQuantXoneTruthPullSnana & \fitPhotozQuantXoneTruthPullJax\\
		$c$ & $\fitPhotozQuantCPullMedSci$ & $\fitPhotozQuantCPullSigSci$ & \fitPhotozQuantCSigRatio & ${\fitPhotozQuantCMadSnana}$ & ${\fitPhotozQuantCMadJax}$ & \fitPhotozQuantCTruthPullSnana & \fitPhotozQuantCTruthPullJax\\
		\hline
	\end{tabular}
\end{table*}

\subsection{Standardisation and cosmological inference}
\label{sec:val_bbc}

We next consider the parity with \snana{}'s standardisation code \textsc{salt2mu} and cosmological inference code \textsc{wfit}, by applying both to 50 simulated realisations of the DES-SN5YR datasets using the simulation inputs from \citet{Vincenzi24, Popovic25}. We assess parity of cosmological results from samples with one fiducial set of analysis choices, a `statistical-only' analysis, and one which explores a variety of different systematic analysis choices, a `systematics analysis'. We consider three variants of this parity test: a statistical-only analysis of simulated spectroscopically-confirmed samples, a statistical-only analysis of simulated photometrically-classified samples including treatment of contaminants, and finally a full systematics analysis of photometrically-classified samples to represent the most realistic case.

These analyses follow DES-SN5YR in using the BBC method summarised in Section~\ref{sec:bbc} \citep{Kessler17}, specifically the BBC-4D approach of \citet{Popovic21} with bias corrections dependent on host-galaxy stellar mass as well as redshift, stretch and colour. This approach requires creating a large simulation many times larger than the observed data to finely model the selection function of the survey, known in \snana{} as a `BIASCOR' simulation. This is a deliberately large simulation to fully quantify the selection bias in the sample across the full parameter space of simulated SN properties. For the photometrically-classified datasets, we use the existing trained \textsc{supernnova} \citep{Moller20} model used within the DES-SN5YR analysis to obtain classification probabilities for each event in the sample. We fit the simulated light curves and apply standardisation, incorporating bias corrections, to construct Hubble diagrams and then infer cosmological parameters for each simulated data sample. We repeat this process using both \snana{} and \codename{} to assess that both codes give consistent results.

In the case of the systematics analysis, different analysis variants are implemented as different options in light curve fitting (known as FITOPTs in \snana{}) and in distance standardisation using \textsc{salt2mu} (known as MUOPTs in \snana{}). Distance estimation is repeated using the full combination of different FITOPTs and MUOPTs, and the resulting shifts in inferred distances across all SNe are used to construct a systematics covariance matrix which describes the covariance between SN distances across the sample. This systematics covariance matrix can then be used within cosmological inference on the sample, yielding cosmological constraints which incorporate uncertainty arising from analysis systematics. For the most complete parity test, we incorporate a large suite of systematic analysis variants as used in the DES-SN5YR analysis \citep{Vincenzi24, Popovic25}, for example considering a variety of different survey cross-calibration solutions \citep{PopovicDovekie}, various forward models of the population of SNe~Ia and permutations of Milky Way extinction. All systematics in the DES-SN5YR analysis are included except for those related to the photometric classifier choice, as \codename{} currently includes only support for one classifier in \textsc{supernnova}, and the systematic associated with peculiar velocities as this is constructed specifically for the real DES-SN5YR sample and is not directly applicable to our simulated sample.

We first consider the results of the statistical-only analyses. Table \ref{tab:bbcparity} presents the difference between recovered $\alpha$, $\beta$ and $\gamma$ Tripp standardisation parameters along with cosmological parameters, as well as the ratio of the uncertainties on each obtained between both codes. Overall both codes produce very consistent results with negligible differences in both inferred parameter values and uncertainties, strongly validating the accuracy of \codename{}.

\begin{table}
	\caption{Median residuals and $\sigma_{\rm NMAD}$ in inferred standardisation and cosmological parameters from \codename{} and \snana{} implementations of \textsc{salt2mu} and \textsc{wfit}, across 50 simulated DES-SN5YR realisations, alongside the median ratio of the uncertainties obtained between the two codes. The parameter scalePCC rescales the classifier's probabilities to correct for a classifier that over- or under-estimates contamination.}
	\label{tab:bbcparity}
	\setlength{\tabcolsep}{4pt}
	\begin{tabular}{lccc}
	\hline
	 & \multicolumn{2}{c}{\codename{} $-$ \snana{} residual} & \\
	Parameter & Median & $\sigma_{\rm NMAD}$ & $\sigma$ ratio\\
	\hline
	\multicolumn{4}{l}{Spectroscopic sample, \bbcParitySpeczNReal{} realisations}\\
	$\alpha$ & $\bbcParitySpeczAlphaDMedSci$ & $\bbcParitySpeczAlphaDMadSci$ & \bbcParitySpeczAlphaSigRatioMed\\
	$\beta$ & $\bbcParitySpeczBetaDMedSci$ & $\bbcParitySpeczBetaDMadSci$ & \bbcParitySpeczBetaSigRatioMed\\
	$\gamma$ & $\bbcParitySpeczGammaDMedSci$ & $\bbcParitySpeczGammaDMadSci$ & \bbcParitySpeczGammaSigRatioMed\\
	$\Omega_{\rm m}$ & $\bbcParitySpeczOmDMedSci$ & $\bbcParitySpeczOmDMadSci$ & \bbcParitySpeczOmSigRatioMed\\
	$w$ & $\bbcParitySpeczWDMedSci$ & $\bbcParitySpeczWDMadSci$ & \bbcParitySpeczWSigRatioMed\\
	\hline
	\multicolumn{4}{l}{Photometric sample, \bbcParityPhotNReal{} realisations}\\
	$\alpha$ & $\bbcParityPhotAlphaDMedSci$ & $\bbcParityPhotAlphaDMadSci$ & \bbcParityPhotAlphaSigRatioMed\\
	$\beta$ & $\bbcParityPhotBetaDMedSci$ & $\bbcParityPhotBetaDMadSci$ & \bbcParityPhotBetaSigRatioMed\\
	$\gamma$ & $\bbcParityPhotGammaDMedSci$ & $\bbcParityPhotGammaDMadSci$ & \bbcParityPhotGammaSigRatioMed\\
	scalePCC & $\bbcParityPhotScalePCCDMedSci$ & $\bbcParityPhotScalePCCDMadSci$ & \bbcParityPhotScalePCCSigRatioMed\\
	$\Omega_{\rm m}$ & $\bbcParityPhotOmDMedSci$ & $\bbcParityPhotOmDMadSci$ & \bbcParityPhotOmSigRatioMed\\
	$w$ & $\bbcParityPhotWDMedSci$ & $\bbcParityPhotWDMadSci$ & \bbcParityPhotWSigRatioMed\\
	\hline
	\end{tabular}
\end{table}

We next consider parity of the two codes when incorporating systematics. We focus on one realisation of the simulated data for this and do not repeat for all 50 simulations; while this is computationally feasible for \codename{}, this would be very expensive for \snana{}. Table \ref{tab:covparity} presents results for all the systematic analysis variants used, corresponding to those in \citet{Vincenzi24} and \citet{Popovic25}; we consider the inferred posterior mean $w$ value and the associated uncertainty $\sigma_{w,\mathrm{syst}}$ arising from each systematic, across both \codename{} and \snana{}. $\sigma_{w,\mathrm{syst}}$ is defined as the additional uncertainty arising on $w$ when each systematic is included when compared to the statistical-only uncertainty: 

\begin{equation}
      \sigma_{w,\mathrm{syst}} = \sqrt{\sigma_{w,\mathrm{tot}}^2-\sigma_{w,\mathrm{stat}}^2}.
  \end{equation}

As demonstrated in Table~\ref{tab:covparity}, we see excellent consistency between the two codes, with any differences of at most $10^{-4}$. Note that in some cases the quoted $\sigma_{w,\mathrm{syst}}$ is 0.0000; this occurs when the individual systematic realisation does not introduce any additional uncertainty on $w$ and $\sigma_{w,\mathrm{tot}}^2-\sigma_{w,\mathrm{stat}}^2$ is negative, meaning there is no real corresponding uncertainty. 

\begin{table}
	\caption{Posterior mean $w$ values and the associated systematic uncertainties, $\sigma_{w,\mathrm{syst}}$, for all systematic analysis variants considered in this work when applied to a simulated DES-SN5YR sample. These are for cosmological constraints when including a prior constraint based on the CMB, as outlined in Section~\ref{sec:val_bbc}.}
	\label{tab:covparity}
	\setlength{\tabcolsep}{4pt}
	\begin{tabular}{lcccc}
	\hline
	 & \multicolumn{2}{c}{$w$} & \multicolumn{2}{c}{$\sigma_{w,\mathrm{syst}}$} \\
	Systematic & \codename{} & \snana{} & \codename{} & \snana{}\\
	\hline
	Statistical-only & $\covParityNosysWJax$ & $\covParityNosysWSnana$ & $\covParityNosysSysJax$ & $\covParityNosysSysSnana$\\
	\hline
	SALT3+Calibration & $\covParityCalSaltWJax$ & $\covParityCalSaltWSnana$ & $\covParityCalSaltSysJax$ & $\covParityCalSaltSysSnana$\\
	HST Calspec & $\covParityCalspecWJax$ & $\covParityCalspecWSnana$ & $\covParityCalspecSysJax$ & $\covParityCalspecSysSnana$\\
	P23 dust pop 1 & $\covParityDustOneWJax$ & $\covParityDustOneWSnana$ & $\covParityDustOneSysJax$ & $\covParityDustOneSysSnana$\\
	P23 dust pop 2 & $\covParityDustTwoWJax$ & $\covParityDustTwoWSnana$ & $\covParityDustTwoSysJax$ & $\covParityDustTwoSysSnana$\\
	P23 dust pop 3 & $\covParityDustThreeWJax$ & $\covParityDustThreeWSnana$ & $\covParityDustThreeSysJax$ & $\covParityDustThreeSysSnana$\\
	BS21 & $\covParityBSTwentyOneWJax$ & $\covParityBSTwentyOneWSnana$ & $\covParityBSTwentyOneSysJax$ & $\covParityBSTwentyOneSysSnana$\\
	P23($u-r$) & $\covParityIntrscColorWJax$ & $\covParityIntrscColorWSnana$ & $\covParityIntrscColorSysJax$ & $\covParityIntrscColorSysSnana$\\
	Model SN~Ia age & $\covParityWTwentyTwoWJax$ & $\covParityWTwentyTwoWSnana$ & $\covParityWTwentyTwoSysJax$ & $\covParityWTwentyTwoSysSnana$\\
	Fixed $\alpha/\beta$ & $\covParityFixabWJax$ & $\covParityFixabWSnana$ & $\covParityFixabSysJax$ & $\covParityFixabSysSnana$\\
	$\alpha$ Evolution & $\covParityAlphaevolWJax$ & $\covParityAlphaevolWSnana$ & $\covParityAlphaevolSysJax$ & $\covParityAlphaevolSysSnana$\\
	$\beta$ Evolution & $\covParityBetaevolWJax$ & $\covParityBetaevolWSnana$ & $\covParityBetaevolSysJax$ & $\covParityBetaevolSysSnana$\\
	$\gamma$ Evolution & $\covParityGammaevolWJax$ & $\covParityGammaevolWSnana$ & $\covParityGammaevolSysJax$ & $\covParityGammaevolSysSnana$\\
	Mass Location & $\covParityMasslocWJax$ & $\covParityMasslocWSnana$ & $\covParityMasslocSysJax$ & $\covParityMasslocSysSnana$\\
	$\sigma_{\rm int}$ modeling & $\covParitySigintModelWJax$ & $\covParitySigintModelWSnana$ & $\covParitySigintModelSysJax$ & $\covParitySigintModelSysSnana$\\
	MW scaling & $\covParityMwebvWJax$ & $\covParityMwebvWSnana$ & $\covParityMwebvSysJax$ & $\covParityMwebvSysSnana$\\
	MW color law & $\covParityColorlawWJax$ & $\covParityColorlawWSnana$ & $\covParityColorlawSysJax$ & $\covParityColorlawSysSnana$\\
	DES SV catalog & $\covParitySvahostWJax$ & $\covParitySvahostWSnana$ & $\covParitySvahostSysJax$ & $\covParitySvahostSysSnana$\\
	Shift in host spec eff & $\covParityEffZspecWJax$ & $\covParityEffZspecWSnana$ & $\covParityEffZspecSysJax$ & $\covParityEffZspecSysSnana$\\
	CC SN prior & $\covParityCcpriorHElevenWJax$ & $\covParityCcpriorHElevenWSnana$ & $\covParityCcpriorHElevenSysJax$ & $\covParityCcpriorHElevenSysSnana$\\
	Redshift shift & $\covParityZshiftWJax$ & $\covParityZshiftWSnana$ & $\covParityZshiftSysJax$ & $\covParityZshiftSysSnana$\\
	\hline
	All systematics & $\covParityAllWJax$ & $\covParityAllWSnana$ & $\covParityAllSysJax$ & $\covParityAllSysSnana$\\
	\hline
	\end{tabular}
\end{table}

We also verify that this full end-to-end cosmological pipeline yields unbiased cosmological constraints by inferring cosmological parameters for each of our 50 realisations, performing both statistical-only and systematics analyses, and assessing whether we recover the truth on average across all of the realisations. We follow previous work \citep[e.g.][]{Kessler23, Vincenzi24, Popovic25} by testing recovery across multiple realisations and incorporating constraints from the CMB. As in these works, we approximate including a full CMB likelihood by applying a prior on the CMB shift parameter $R$, as defined in \citet{Komatsu09}. The prior is centred around the true simulated cosmological parameters with a prior uncertainty of $\sigma_R=0.006$ as in \citet{Kessler23}, to represent constraints from Planck \citep{Planck20}. Across the 50 realisations, the mean offsets from the simulated cosmology are $\Delta w = +0.0008 \pm 0.0021$ and $\Delta\Omega_{\rm m} = -0.0003 \pm 0.0006$ for the spectroscopically-confirmed sample, consistent with zero. For the photometrically-classified sample, there are small biases of $\Delta w = +0.0084 \pm 0.0020$ and $\Delta\Omega_{\rm m} = +0.0012 \pm 0.0006$ but these are consistent with what \snana{} also finds and negligible compared to the uncertainty on $w$ from a single realisation ($\sigma_{w,\text{stat}} \approx 0.022$). Including systematics, the mean offsets are $\Delta w = -0.0075 \pm 0.0029$ and $\Delta\Omega_{\rm m} = -0.0019 \pm 0.0008$, again small compared to the full uncertainty on $w$ from a single realisation ($\sigma_{w,\text{stat+syst}} \approx 0.031$). \codename{} is therefore demonstrated to yield unbiased cosmological constraints in a full \jax{} end-to-end pipeline. 

\section{Performance}
\label{sec:performance}

Having established that \codename{} provides robust results across a variety of different tasks, we now present benchmarks detailing performance of simulations and light curve fitting on two types of hardware to represent the experience of different users; a CPU benchmark for an Apple MacBook Pro (with M1 chip) running a CPU-only version of \jax{} to represent a user running on their own laptop, and a GPU benchmark on an NVIDIA A100 to represent the experience of a user running on an HPC cluster.

We consider a few different performance benchmarks, spanning both simulation and light curve fitting:

\begin{enumerate}
    \item We express two separate performance metrics, end-to-end run time and the post-JIT rate. The former includes the full runtime of the script from start to finish, from imports all the way to saving the output files, representative of real use. The latter is the rate of simulation and fitting post-JIT, of relevance for an SBI application where the same methods are called repeatedly to feed into a neural network and outputs are kept in memory rather than saved to disk. JIT-compilation makes up a significant portion of the runtime; for larger runs, the cost of the compilation is amortised over a larger total number of events leading to an overall improvement in efficiency. The post-JIT rate is therefore reflective of the maximum level of performance.
    \item We also consider the impact of numerical precision, exploring both single-precision (32 bit) and double-precision (64 bit) floats. \snana{} runs in double-precision and \codename{} supports double-precision, but some GPUs have fewer cores for double-precision arithmetic leading to reduced performance; as a result \codename{} supports both options. Both give good parity with \snana{}, and we consider the performance impact of working at both precision levels.
    \item When considering LSST simulations we present simulations of the DDF and Wide Fast Deep (WFD) survey separately \citep{Ivezic19}. DDFs are observed at a much higher cadence than WFD and \codename{} involves light curve padding to the longest sequence to create a standard grid shape, as discussed in Section~\ref{sec:design}. As a result, DDF and WFD have been separated to avoid WFD simulations being padded to the length of DDF light curves, which would increase computational overhead. We present both results to illustrate performance with different types of simulation.
    \item Comparing numbers for a GPU with those for a single CPU core is somewhat misleading since a GPU hour is more expensive than a CPU hour. To present a fairer comparison, we additionally present a representative number for \snana{} benchmarks for 32 CPU cores, which is expressed as an approximate value and obtained by multiplying the single core number by 32 rather than through a separate benchmark. For compute nodes on HPC resources such as the Cambridge Service for Data Driven Discovery (CSD3) and NERSC Perlmutter, CPU nodes typically have approximately 128 CPU cores and GPU nodes have 4 GPUs, and using cloud computing platforms one A100 GPU hour roughly corresponds to 32 CPU hours in cost\footnote{At the time of writing, on Google Cloud a CPU core costs \$0.097 per hour while an A100 GPU is \$3.67 per hour, making an A100 GPU equivalent to 38 CPU cores in cost. We use 32 as a representative number roughly corresponding to both cloud compute platforms and HPC resources.}. These numbers enable an illustrative comparison between CPU and GPU numbers, though are quoted as approximate as they will depend on the CPU in question.
\end{enumerate} 

Quoted numbers compare the performance obtained running \snana{} and \codename{} with the same input files. While we include the 32 core number to represent comparable performance costs, the highest absolute throughput can always be obtained by running on GPU; if the goal is speed, running on 10 GPUs would give an additional order-of-magnitude performance improvement as many of these tasks are embarrassingly parallel.

\subsection{Supernova cosmology}

We first consider performance of the various elements in the supernova cosmology pipeline, where the slowest tasks are light curve fitting and then simulation. We consider simulations and fits with both SALT and BayeSN, as well as the performance of the full end-to-end pipeline.

\subsubsection{CPU}
\label{sec:perf_cpu}

We first consider performance on CPU, comparing \snana{} to \codename{} for a variety of SALT and BayeSN simulations and fits, along with photo-$z$ fits for SALT and BayeSN. While the main goal of \codename{} has been to leverage GPU-acceleration, it is still very performant when running on CPU; this is illustrated in Table \ref{tab:performance}, which presents performance as measured by throughput, number of SNe/sec, for a variety of tasks.

As discussed in Section~\ref{sec:performance}, we present both end-to-end and post-JIT throughputs. Compilation makes up a significant proportion of the runtime, therefore running larger jobs simulating and fitting more supernovae amortises the cost of compilation over a larger total number of events. As the total number of events increases, the end-to-end performance will increase and approach the post-JIT rate. Quoted CPU benchmarks are based on simulating samples of \benchSimSaltN{} events and fitting samples of \benchFitSaltN{} (spec-$z$) and \benchFitpzSaltN{} (photo-$z$) events.

For brevity we will not repeat every number from the table in the text and focus on the overall changes between the different approaches. For end-to-end simulation time, \snana{} simulates between 100-700 SNe/sec depending on the configuration; \codename{} produces an order-of-magnitude increase of 1,500-7,900 SNe/sec end-to-end, with post-JIT rates as high as 12,000 SNe/sec for SALT simulations in single precision. We see a similar order-of-magnitude increase for light curve fitting, both with a spec-$z$ and a photo-$z$, with further speed-ups post-JIT. 

It should be noted that \codename{} runs a fixed number of optimiser iterations for every light curve (25 in the first fitting round and 15 in the second and third), as discussed in Section~\ref{sec:fit}, whereas \snana{} iterates each fit until convergence. Additionally, \codename{} will utilise multiple cores when running on CPU, while \snana{} processes are restricted to a single core, though multiple \snana{} processes can be run in parallel. It is not straightforward to express \codename{} performance per core, as JAX will dynamically use a different number of cores depending on the specific structure of the compiled code being run. \codename{} typically uses around 2-6 CPU cores while running these benchmarks, varying by task. Even considering this, \codename{} shows a significant performance increase for many tasks, and the ability to naturally use multiple cores to boost performance dynamically rather than requiring multiple independent processes to be run is an advantage.

\subsubsection{GPU}
\label{sec:perf_gpu}

We next consider performance for the same tasks as in Section~\ref{sec:perf_cpu}, except on an NVIDIA A100 GPU. Table \ref{tab:performance} again illustrates the performance, showcased in terms of throughput of number of SNe/sec. There is a very notable performance increase when switching from CPU to GPU, another order of magnitude. The headline results are that end-to-end simulations reach 13-16,000 SNe/sec for LSST DDF simulations and exceed 20,000 SNe/sec for WFD fields, while post-JIT performance exceeds 100,000 SNe/sec across both single- and double-precision in WFD. Compared to the representative 32 core \snana{} numbers, simulations on GPU are approximately an order of magnitude faster post-JIT.

For light curve fitting, post-JIT \codename{} is able to fit thousands of SN light curves per second, 5,000 SNe/sec for SALT3 and 6,200 SNe/sec for BayeSN in single precision, with values roughly halved in double precision. Comparing to the 32 core \snana{} values, the improvement for light curve fitting is notably larger than for simulations, with a 30-50 times increase post-JIT and a 14-32 times increase end-to-end in single precision. Overall, running light curve fits using \codename{} on GPU provides a step-change in computational performance.

\begin{table*}
	\centering
	\caption{Performance metrics for \snana{} and \codename{} across a variety of simulation and fitting tasks using SALT and BayeSN models, measured on both CPU and GPU for the latter, expressed in terms of throughput (SNe/sec). Columns give the end-to-end (e2e) throughput and the post-JIT rate, after compilation. Sample sizes $N$ (CPU, GPU): simulation \benchSimSaltN, \benchSimSaltGpuN; spec-$z$ fitting \benchFitSaltN, \benchFitSaltGpuN; photo-$z$ fitting \benchFitpzSaltN, \benchFitpzSaltGpuN. The 32 core \snana{} numbers are approximate, obtained by multiplying the single core \snana{} numbers by 32, to provide a representative comparison to GPU performance, in terms of computational expense, as discussed in Section~\ref{sec:performance}.}
	\label{tab:performance}
	\setlength{\tabcolsep}{4pt}
	\begin{tabular}{lrrrrrrrrrr}
		\hline
		 & \multicolumn{5}{c}{CPU} & \benchSnanaCoresPerGpu{} cores & \multicolumn{4}{c}{GPU} \\
		 & \snana{} & \multicolumn{4}{c}{\codename{}} & \snana{} & \multicolumn{4}{c}{\codename{}} \\
		 & double & \multicolumn{2}{c}{single} & \multicolumn{2}{c}{double} & double & \multicolumn{2}{c}{single} & \multicolumn{2}{c}{double} \\
		Task & & e2e & post-JIT & e2e & post-JIT & & e2e & post-JIT & e2e & post-JIT\\
		\hline
		\multicolumn{11}{l}{\textit{Simulation, LSST DDF}}\\
		\quad \salt{}3 & \benchSimSaltDdfSnana & \benchSimSaltDdfCpuSingle & \benchSimSaltDdfCpuSingleSteady & \benchSimSaltDdfCpuDouble & \benchSimSaltDdfCpuDoubleSteady & $\approx$\benchSimSaltDdfSnanaCores & \benchSimSaltDdfGpuSingle & \benchSimSaltDdfGpuSingleSteady & \benchSimSaltDdfGpuDouble & \benchSimSaltDdfGpuDoubleSteady\\
		\quad \bayesn{} & \benchSimBayesnDdfSnana & \benchSimBayesnDdfCpuSingle & \benchSimBayesnDdfCpuSingleSteady & \benchSimBayesnDdfCpuDouble & \benchSimBayesnDdfCpuDoubleSteady & $\approx$\benchSimBayesnDdfSnanaCores & \benchSimBayesnDdfGpuSingle & \benchSimBayesnDdfGpuSingleSteady & \benchSimBayesnDdfGpuDouble & \benchSimBayesnDdfGpuDoubleSteady\\
		\multicolumn{11}{l}{\textit{Simulation, LSST WFD}}\\
		\quad \salt{}3 & \benchSimSaltWfdSnana & \benchSimSaltWfdCpuSingle & \benchSimSaltWfdCpuSingleSteady & \benchSimSaltWfdCpuDouble & \benchSimSaltWfdCpuDoubleSteady & $\approx$\benchSimSaltWfdSnanaCores & \benchSimSaltWfdGpuSingle & \benchSimSaltWfdGpuSingleSteady & \benchSimSaltWfdGpuDouble & \benchSimSaltWfdGpuDoubleSteady\\
		\quad \bayesn{} & \benchSimBayesnWfdSnana & \benchSimBayesnWfdCpuSingle & \benchSimBayesnWfdCpuSingleSteady & \benchSimBayesnWfdCpuDouble & \benchSimBayesnWfdCpuDoubleSteady & $\approx$\benchSimBayesnWfdSnanaCores & \benchSimBayesnWfdGpuSingle & \benchSimBayesnWfdGpuSingleSteady & \benchSimBayesnWfdGpuDouble & \benchSimBayesnWfdGpuDoubleSteady\\
		\multicolumn{11}{l}{\textit{Fitting, spectroscopic redshift, DES-SN5YR simulations}}\\
		\quad \salt{}3 & \benchFitSaltSnana & \benchFitSaltCpuSingle & \benchFitSaltCpuSingleSteady & \benchFitSaltCpuDouble & \benchFitSaltCpuDoubleSteady & $\approx$\benchFitSaltSnanaCores & \benchFitSaltGpuSingle & \benchFitSaltGpuSingleSteady & \benchFitSaltGpuDouble & \benchFitSaltGpuDoubleSteady\\
		\quad \bayesn{} & -- & \benchFitBayesnCpuSingle & \benchFitBayesnCpuSingleSteady & \benchFitBayesnCpuDouble & \benchFitBayesnCpuDoubleSteady & -- & \benchFitBayesnGpuSingle & \benchFitBayesnGpuSingleSteady & \benchFitBayesnGpuDouble & \benchFitBayesnGpuDoubleSteady\\
		\multicolumn{11}{l}{\textit{Fitting, photometric redshift, LSST DDF}}\\
		\quad \salt{}3 & \benchFitpzSaltDdfSnana & \benchFitpzSaltDdfCpuSingle & \benchFitpzSaltDdfCpuSingleSteady & \benchFitpzSaltDdfCpuDouble & \benchFitpzSaltDdfCpuDoubleSteady & $\approx$\benchFitpzSaltDdfSnanaCores & \benchFitpzSaltDdfGpuSingle & \benchFitpzSaltDdfGpuSingleSteady & \benchFitpzSaltDdfGpuDouble & \benchFitpzSaltDdfGpuDoubleSteady\\
		\quad \bayesn{} & -- & \benchFitpzBayesnDdfCpuSingle & \benchFitpzBayesnDdfCpuSingleSteady & \benchFitpzBayesnDdfCpuDouble & \benchFitpzBayesnDdfCpuDoubleSteady & -- & \benchFitpzBayesnDdfGpuSingle & \benchFitpzBayesnDdfGpuSingleSteady & \benchFitpzBayesnDdfGpuDouble & \benchFitpzBayesnDdfGpuDoubleSteady\\
		\multicolumn{11}{l}{\textit{Fitting, photometric redshift, LSST WFD}}\\
		\quad \salt{}3 & \benchFitpzSaltWfdSnana & \benchFitpzSaltWfdCpuSingle & \benchFitpzSaltWfdCpuSingleSteady & \benchFitpzSaltWfdCpuDouble & \benchFitpzSaltWfdCpuDoubleSteady & $\approx$\benchFitpzSaltWfdSnanaCores & \benchFitpzSaltWfdGpuSingle & \benchFitpzSaltWfdGpuSingleSteady & \benchFitpzSaltWfdGpuDouble & \benchFitpzSaltWfdGpuDoubleSteady\\
		\quad \bayesn{} & -- & \benchFitpzBayesnWfdCpuSingle & \benchFitpzBayesnWfdCpuSingleSteady & \benchFitpzBayesnWfdCpuDouble & \benchFitpzBayesnWfdCpuDoubleSteady & -- & \benchFitpzBayesnWfdGpuSingle & \benchFitpzBayesnWfdGpuSingleSteady & \benchFitpzBayesnWfdGpuDouble & \benchFitpzBayesnWfdGpuDoubleSteady\\
		\hline
	\end{tabular}
\end{table*}

\subsubsection{Full cosmological pipeline}

As presented in Section~\ref{sec:val_bbc}, within this work we test the full end-to-end BBC cosmology pipeline with the same inputs as used for the DES-SN5YR analysis. For a statistical-only analysis, this involves simulating approximately 100 million light curves, fitting nearly 4 million light curves (many of the simulations are discarded by selection cuts), standardising 50 realised datasets and then inferring cosmological parameters for each of them. This very large scale task was all completed within a single \textsc{python} script in less than 3 hours running on a single GPU. We additionally repeat this for a systematics analysis with 50 data realisations, which involves repeating many of these stages with different analysis variations, approximately 750 million simulated light curves and 30 million light curve fits alongside construction of systematics covariance matrices and cosmological inference for the sample. This took less than 20 hours on a single GPU. Overall, these results serve to showcase the considerable performance benefits of \codename{}.

\subsection{\textsc{redback-jax} models}

We next consider performance of physical models from \textsc{redback-jax} (\textcolor{blue}{Sarin et al., in prep.}), focusing on the Arnett model of radioactive decay for light curve simulations and fitting. These results are presented in Table~\ref{tab:performance_arnett}, based on samples of \benchSimArnettN{} simulated and \benchFitArnettN{} fitted events on CPU, and \benchSimArnettGpuN{} simulated and \benchFitArnettGpuN{} fitted events on GPU. Performance for simulation is comparable to SALT and BayeSN simulations; 800-1,300 SNe/sec end-to-end on CPU, 2,400-6,800 SNe/sec post-JIT depending on the precision and cadence. 

We also consider light curve fits. These use the same combination of priors as used for the fitting results presented in Section~\ref{sec:arnett_fitting}, fitting for the explosion time, $M_{\rm ej}$, $f_{\rm Ni}$, $v_{\rm ej}$ and $\kappa$. Performance is fast, although not quite as fast as SALT or BayeSN fits; fits run at 260 SNe/sec end-to-end in single-precision and 130 SNe/sec in double precision. For these fits, the end-to-end performance is far closer to the post-JIT performance as the fits are generally more expensive, and therefore compilation makes up a lower proportion of the total run times.

\begin{table*}
	\centering
	\caption{Performance metrics of \codename{} for Arnett model simulation and light curve fitting tasks,  measured on both CPU and GPU, expressed in terms of throughput (SNe/sec). Columns give the end-to-end (e2e) throughput and the post-JIT rate, after compilation. Sample sizes $N$ (CPU, GPU): simulation \benchSimArnettN, \benchSimArnettGpuN; fitting \benchFitArnettN, \benchFitArnettGpuN.}
	\label{tab:performance_arnett}
	\setlength{\tabcolsep}{4pt}
	\begin{tabular}{lrrrrrrrr}
		\hline
		 & \multicolumn{4}{c}{CPU} & \multicolumn{4}{c}{GPU} \\
		 & \multicolumn{2}{c}{single} & \multicolumn{2}{c}{double} & \multicolumn{2}{c}{single} & \multicolumn{2}{c}{double} \\
		Task & e2e & post-JIT & e2e & post-JIT & e2e & post-JIT & e2e & post-JIT\\
		\hline
		Simulation, LSST DDF & \benchSimArnettDdfCpuSingle & \benchSimArnettDdfCpuSingleSteady & \benchSimArnettDdfCpuDouble & \benchSimArnettDdfCpuDoubleSteady & \benchSimArnettDdfGpuSingle & \benchSimArnettDdfGpuSingleSteady & \benchSimArnettDdfGpuDouble & \benchSimArnettDdfGpuDoubleSteady\\
		Simulation, LSST WFD & \benchSimArnettWfdCpuSingle & \benchSimArnettWfdCpuSingleSteady & \benchSimArnettWfdCpuDouble & \benchSimArnettWfdCpuDoubleSteady & \benchSimArnettWfdGpuSingle & \benchSimArnettWfdGpuSingleSteady & \benchSimArnettWfdGpuDouble & \benchSimArnettWfdGpuDoubleSteady\\
		Fitting, LSST DDF & \benchFitArnettDdfCpuSingle & \benchFitArnettDdfCpuSingleSteady & \benchFitArnettDdfCpuDouble & \benchFitArnettDdfCpuDoubleSteady & \benchFitArnettDdfGpuSingle & \benchFitArnettDdfGpuSingleSteady & \benchFitArnettDdfGpuDouble & \benchFitArnettDdfGpuDoubleSteady\\
		Fitting, LSST WFD & \benchFitArnettWfdCpuSingle & \benchFitArnettWfdCpuSingleSteady & \benchFitArnettWfdCpuDouble & \benchFitArnettWfdCpuDoubleSteady & \benchFitArnettWfdGpuSingle & \benchFitArnettWfdGpuSingleSteady & \benchFitArnettWfdGpuDouble & \benchFitArnettWfdGpuDoubleSteady\\
		\hline
	\end{tabular}
\end{table*}

\section{Discussion}
\label{sec:discussion}

Through new surveys such as LSST \citep{Ivezic19} and Roman's High-Latitude Time Domain Survey \citep{Rose21, Kessler25}, supernova cosmology and transient astronomy face both unprecedented challenges and opportunities. The vast scale of upcoming datasets has strongly motivated the development of new analysis techniques \citep[e.g.][]{Rubin25, Karchev23a, Karchev23b, Karchev24Side, Boyd26, Karchev26, Popovic26} which are suitable to be used at these scales. As discussed in Sections~\ref{sec:validation} and~\ref{sec:performance}, \codename{} provides robust simulations, light curve fitting, standardisation and cosmological inference with transformational speedups over existing tools; two orders of magnitude faster than \snana. 

This level of scalability is essential considering that LSST is set to discover upwards of a hundred thousand SNe~Ia \citep{DESC18SRD} over the next decade. The BBC cosmological pipeline used for DES-SN5YR and SN UNITE \citep{Camilleri26, Lee26} relies on large BIASCOR simulations, many times larger than observed datasets, to measure in fine detail the selection bias associated with a survey with good representation across the full parameter space of SN properties. For DES-SN5YR, this involved simulating and fitting samples of millions of events for a sample of approximately 2,000 events. At LSST scale, the vast simulations required to quantify selection bias will be prohibitively expensive using existing tools. For DES-SN5YR the BIASCOR simulation is $\sim$1,000 times larger than the real dataset; applying this to an anticipated LSST sample of order 100,000 SNe, a BIASCOR simulation would need to be of order 100 million simulated SNe, and many such simulations would need to be created for a full systematics analysis. One BIASCOR simulation would take 8,200 CPU core hours to fit with SALT using \snana{}, compared to 5 GPU hours in single-precision using \codename{}; this performance speed-up is vital. 

Of course, other methods are being developed for cosmological inference which may in time supersede the BBC pipeline. However, many of these are SBI methods which themselves rely on the ability to simulate, and sometimes fit, large numbers of different datasets; a performant GPU-accelerated simulator is also highly desirable to train SBI methods. \codename{} has been specifically designed to address this need, allowing users to create fully compiled simulators and use them to create on-the-fly simulations within the training loops of SBI training algorithms. Future work will present SBI analyses which use \codename{} to infer population-level properties of transient populations incorporating selection effects.

One of the key strengths of \codename{} is its accessibility and its modular, extensible design. As described in Section~\ref{sec:sn_host}, users can specify relations between SN and host-galaxy properties directly, and draw SNe from sub-populations whose relative rates depend on the host, to explore the impact of different astrophysical relationships on observations. This same modular design allows users to add new SED models and photometric classifiers and create custom analysis pipelines.

Overall, through high performance, accessibility, modularity and extensibility, \codename{} addresses many of the key challenges currently faced in supernova cosmology.

\section{Conclusions}
\label{sec:conclusions}

In this work we present \codename{}, a \jax{} implementation of the \snana{} framework with order-of-magnitude performance increases through support for GPU acceleration. Our key findings and notable features of \codename{} are summarised as follows:

\begin{enumerate}
    \item \codename{} is a comprehensive framework for supernova cosmology and transient astronomy, supporting realistic survey simulations including both spectra and photometry, light curve fitting, standardisation and cosmological inference. It currently builds in support for a variety of SN models: the SN~Ia SED models SALT and BayeSN, various transient SED templates for core-collapse SNe and peculiar SNe~Ia \citep{Jones17, Vincenzi19, Kessler19PLAsTiCC}, and physical models from \textsc{redback-jax}. The code is designed to be modular and extensible to allow users to incorporate new models and features.
    \item \codename{} allows the user to simulate generic customisable relationships between SN properties and host galaxy properties, and to simulate multiple populations of transients linked to different host galaxy properties.
    \item \codename{} has been specifically designed to support future SBI analyses of transient populations, allowing users to simulate over a hundred thousand events per second and use a JIT-compiled simulator to create on-the-fly simulations for training downstream SBI methods.
    \item We validate the accuracy of \codename{} across a variety of tasks, finding that it gives robust results when cross-comparing with outputs from \snana{}; mmag-level agreement in simulated photometry, very similar results from light curve fitting and excellent consistency in standardisation and cosmological parameters. Overall \codename{} demonstrates close parity with \snana{}.
    \item We assess the computational performance of \codename{}, finding a transformational performance increase. When running on CPU performance is increased more than an order of magnitude, and by more than two orders of magnitude on GPU. While the exact level of performance will depend on the level of floating-point precision and the model being used, \codename{} is able to simulate more than a hundred-thousand SNe/sec on GPU post-JIT and fit 5,000 SNe/sec. Running a full end-to-end cosmological analysis on 50 simulated DES-SN5YR samples, with a full treatment of systematics spanning 750 million light curve simulations and 30 million fits as well as standardisation and cosmological inference, runs in under 20 hours on a single GPU. \codename{} is ideally suited for the large data samples expected from LSST and Roman.
    \item \codename{} can be used to simulate hundreds-of-thousands of SNe per second using \textsc{redback-jax} models, including the Arnett model \citep{Arnett82, Pinto00, Sarin26b} and magnetar models \citep{Nicholl17, Guillochon18, Omand24, Sarin22}, as well as fitting hundreds of SNe per second with these models. 
    \item \codename{} is now publicly-available at \url{https://github.com/mattgrayling/jax-snana} and pip-installable. It has been designed to be modular and accessible, allowing the community to extend it with new functionality and develop their own analysis pipelines.
\end{enumerate}

Considering its high level of performance, extensibility and wide variety of functionality, \codename{} has the potential to play a major role in supernova cosmology and transient astronomy over the coming years.

\section*{Acknowledgements}

This work was supported in part by the European Union's Horizon 2020 research and innovation programme under European Research Council Grant Agreement No. 101002652 (BayeSN; PI K. Mandel). MG was supported by the research environment and infrastructure of the Handley Lab at the University of Cambridge. MG would like to acknowledge the expertise and endless dedication of Rick Kessler, who developed \snana{} and never fails to make himself available to provide support to those wishing to use it. MG also thanks Brodie Popovic, Rebecca Chen, Nikhil Sarin and Tamara Davis for valuable discussions about this work, and Will Handley for the continued support and inspiration for this project. 

\section*{Data Availability}

The \codename{} code is publicly available at \url{https://github.com/mattgrayling/jax-snana}. This work makes use of the public \textsc{SNDATA\_ROOT} repository (\url{https://doi.org/10.5281/zenodo.19503606}), containing many of the key files used by \snana{}, and input files from the DES-SN5YR data release (\url{https://doi.org/10.5281/zenodo.12720778}).

\section*{AI Disclosure Statement}

The development of \codename{} made extensive use of agentic coding tools, Claude Code and OpenAI's Codex (across a variety of different Sonnet, Opus, Fable, Astra and Sol model versions), which were also used to create a number of analysis scripts for this work. However, the manuscript was written by the author with LLM usage restricted to copy-editing.


\bibliographystyle{mnras}
\bibliography{refs}


\appendix

\section{Photometric-redshift fits with Gaussian and no host prior}
\label{app:photoz_priors}

\begin{figure*}
	\includegraphics[width=\textwidth]{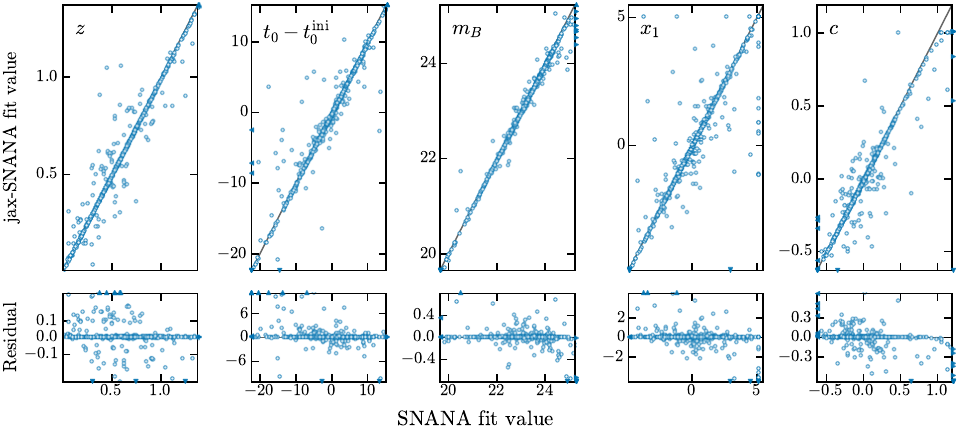}
    \caption{\textbf{Top}: Comparison of fitted values for SALT fits, with Gaussian photo-$z$ priors and redshift a fitted parameter alongside SALT parameters, to the same simulated SNe using both \codename{} and SNANA to establish parity between the two. \textbf{Bottom}: Residuals of the top panels.}
    \label{fig:fit_parity_photoz_gauss}
\end{figure*}

\begin{figure*}
	\includegraphics[width=\textwidth]{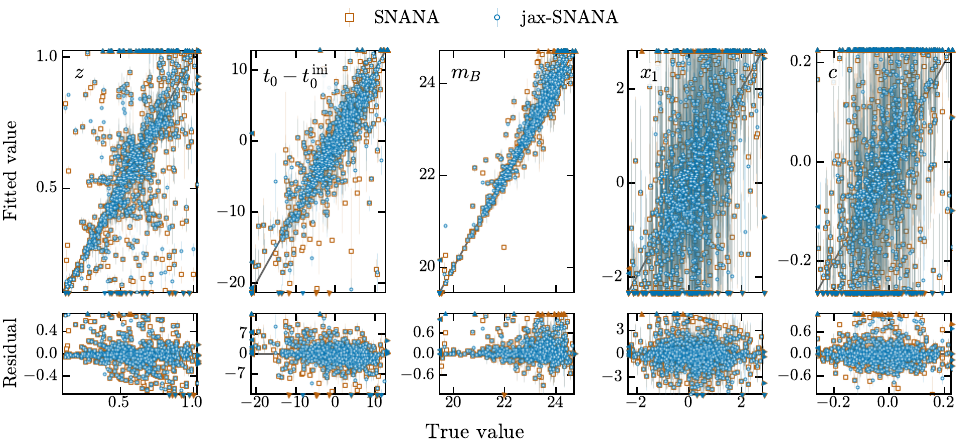}
    \caption{\textbf{Top}: Comparison between fitted values and truth for SALT fits with Gaussian photo-$z$ priors, to the same simulated SNe presented in Fig.~\ref{fig:fit_parity_photoz_gauss}, using both \codename{} and SNANA. \textbf{Bottom}: Residuals of the top panels.}
    \label{fig:fit_recovery_photoz_gauss}
\end{figure*}

\begin{figure*}
	\includegraphics[width=\textwidth]{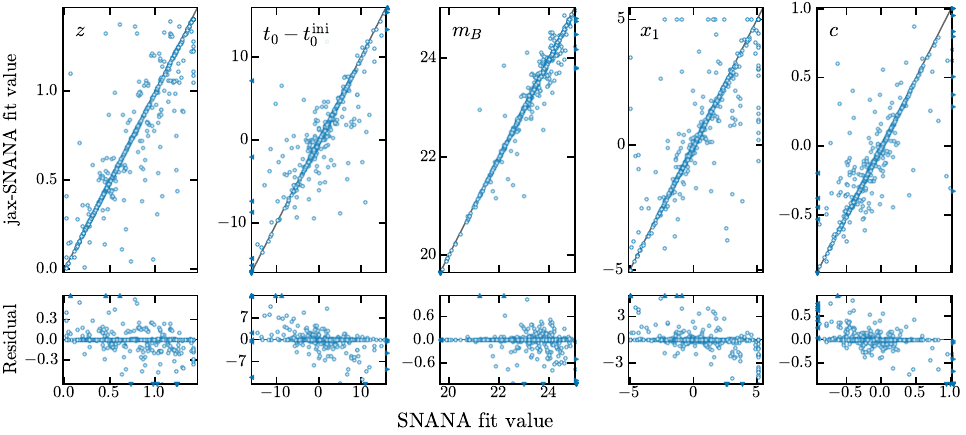}
    \caption{\textbf{Top}: Comparison of fitted values for SALT fits, with no photo-$z$ prior and redshift a fitted parameter alongside SALT parameters, to the same simulated SNe using both \codename{} and SNANA to establish parity between the two. \textbf{Bottom}: Residuals of the top panels.}
    \label{fig:fit_parity_photoz_noprior}
\end{figure*}

\begin{figure*}
	\includegraphics[width=\textwidth]{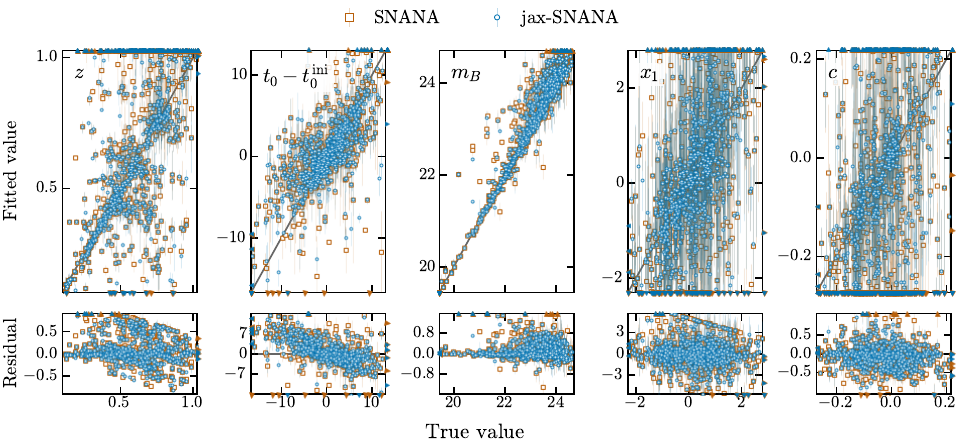}
    \caption{\textbf{Top}: Comparison between fitted values and truth for SALT fits with no photo-$z$ prior, to the same simulated SNe presented in Fig.~\ref{fig:fit_parity_photoz_noprior}, using both \codename{} and SNANA. \textbf{Bottom}: Residuals of the top panels.}
    \label{fig:fit_recovery_photoz_noprior}
\end{figure*}

Figure \ref{fig:fit_parity_photoz_gauss} presents the residual between fitted parameters obtained using \codename{} compared to SNANA for light curve fits with a Gaussian photo-$z$ prior, while Fig.~\ref{fig:fit_recovery_photoz_gauss} presents fitted values compared to true values for both codes, showing an overall high level of consistency albeit with some outliers where both approaches converged to different local minima. Figures~\ref{fig:fit_parity_photoz_noprior} and \ref{fig:fit_recovery_photoz_noprior} are the corresponding plots for the light curve fits without a redshift prior.

\bsp	
\label{lastpage}
\end{document}